\documentclass{aa}
\usepackage{lscape}
\usepackage{subcaption}
\bibpunct{(}{)}{;}{a}{}{,}

\title{X-ray spectra of the Fe-L complex III: systematic uncertainties in the atomic data}
\author{Liyi Gu \inst{1,2,10} 
\and 
Chintan Shah \inst{3,4}
\and 
Junjie Mao \inst{5,6,1}
\and 
A.J.J. Raassen \inst{1,7}
\and
Jelle de Plaa \inst{1}
\and 
Ciro Pinto \inst{8}
\and 
Hiroki Akamatsu \inst{1} 
\and 
Norbert Werner \inst{9,5}
\and 
Aurora Simionescu \inst{1,10,11}
\and 
François Mernier \inst{12,1}
\and 
Makoto Sawada \inst{2}
\and
Pranav Mohanty \inst{10}
\and 
Pedro Amaro \inst{13}
\and 
Ming Feng Gu \inst{14}
\and 
F. Scott Porter \inst{3} 
\and 
José R. Crespo López-Urrutia \inst{4}
\and 
Jelle S. Kaastra \inst{1,10} 
}
\institute{
SRON Netherlands Institute for Space Research, Niels Bohrweg 4, 2333 CA Leiden, the Netherlands
\and
RIKEN High Energy Astrophysics Laboratory, 2-1 Hirosawa, Wako, Saitama 351-0198, Japan
\and 
NASA/Goddard Space Flight Center, 8800 Greenbelt Rd, Greenbelt, MD 20771, USA
\and 
Max-Planck-Institut f$\rm \ddot{u}$r Kernphysik, Heidelberg, D-69117 Heidelberg, Germany
\and 
Department of Physics, Hiroshima University, 1-3-1 Kagamiyama, Higashi-Hiroshima, Hiroshima 739-8526, Japan
\and
Department of Physics, University of Strathclyde, Glasgow, G4 0NG, UK
\and 
Astronomical Institute ``Anton Pannekoek'', Science Park 904, 1098 XH Amsterdam, University of Amsterdam, The Netherlands
\and 
INAF - IASF Palermo, Via U. La Malfa 153, 90146 Palermo PA, Italy
\and 
Department of Theoretical Physics and Astrophysics, Faculty of Science, Masaryk University, Kotl$\rm \acute{a}$$\rm \breve{r}$sk$\rm \acute{a}$ 2, Brno, 611 37, Czech Republic
\and 
Leiden Observatory, Leiden University, PO Box 9513, 2300 RA Leiden, the Netherlands 
\and
Kavli Institute for the Physics and Mathematics of the Universe (WPI), University of Tokyo, Kashiwa 277-8583, Japan  
\and 
ESTEC/ESA, Keplerlaan 1, 2201AZ Noordwijk, The Netherlands
\and 
Laboratory of Instrumentation, Biomedical Engineering and Radiation Physics (LIBPhys-UNL), Department of Physics, NOVA School of Science and Technology, NOVA University Lisbon, 2829-516 Caparica, Portugal
\and 
Space Science Laboratory, University of California, Berkeley, CA 94720, USA
}

\abstract{ There has been a growing request from the X-ray astronomy community for a quantitative estimate of systematic uncertainties originating from the atomic data used in plasma codes.
Though there have been several studies looking into atomic data uncertainties using theoretical calculations, 
in general, there is no commonly accepted solution for this task.
We present a new approach for estimating uncertainties in the line emissivities for the current models of collisional plasma, 
mainly based upon dedicated analysis of 
observed high resolution spectra of stellar coronae and galaxy clusters. We find that the systematic uncertainties of the observed lines consistently show 
anti-correlation with the model line fluxes, after properly accounting for the additional uncertainties from the ion concentration calculation. The strong lines in the spectra are in general better reproduced, indicating that the atomic data and modeling of the main transitions are more accurate than those for the minor ones. This underlying anti-correlation
is found to be roughly independent on source properties, line positions, ion species, and the line formation processes. We further apply our method to the simulated 
XRISM and Athena
observations of collisional plasma sources and discuss the impact of uncertainties on the interpretation of these spectra. The typical uncertainties are $1-2$\% on temperature and $3-20$\% on abundances of O, Ne, Fe, Mg, and Ni. 

}

\titlerunning{Fe-L complex part III }
\authorrunning{L. Gu}

\begin{document}

\maketitle

\section{Introduction}

Studies of astrophysical sources involve analyses of spectra for diagnostics of plasma parameters: 
density, temperature, ionization states, chemical composition, dynamics, and the underlying energy source. Interpretation
of astrophysical spectra requires a huge atomic database including data such as 
energy levels, transition probabilities, excitation rates,
and the ionization balance of plasma. Most of these data are obtained in theoretical calculations with only a few benchmarks performed
with laboratory measurements \citep{kaastra1996, dere1997, smith2001, foster2012, dp2019, dz2020}. There is an increasing demand by the astronomical community that the plasma modeling should include an uncertainty
estimate alongside the numerical values provided, which are essential for astronomers to know how the atomic data could affect 
the accuracy of their final results obtained from the spectra.

Despite the obvious need, so far there is no straightforward way to assess the uncertainties of atomic data and how 
these propagate into the spectral parameters. Recently, a growing effort has been made: there are several
studies attempting to define uncertainties on existing atomic data based on the spread of different fundamental theoretical 
calculations \citep{bautista2013, loch2013, mehdipour2016, yu2018, hitomiatomic}, or from perturbation terms in solving the many-body 
Schrödinger equation \citep{chung2016}. These uncertainties can be propagated through plasma modeling using a quasi-analytic
algorithm \citep{bautista2013}, a Monte-Carlo method \citep{loch2013, hitomiatomic, dz2019}, or affect the interpretation of observational data
with a Bayesian approach \citep{yu2018}.

These theoretical approaches often require sampling of a large amount of relevant level and transition calculations, taking into 
account complexities such as correlated uncertainties in different transitions. This makes most of the theoretical 
approaches too computational demanding for practical analysis of the observed spectra. 
Here we propose an alternative solution. The model
uncertainties could in principle be inferred by comparing with the real data, for instance, through a statistical sampling of discrepancies between theoretical models and
well-calibrated, high-quality spectra taken from laboratory measurements and/or observations, which can be regarded as the absolute true 
values within their quoted uncertainties. The observational constraint on model uncertainties 
will be useful if (1) the sampling size of spectral features (e.g., emission lines) is 
statistically significant, and (2) the observed discrepancies
are not driven by other types of uncertainties, e.g., statistical uncertainty and 
systematic uncertainty from instrumental calibration. Although this approach may not explore the scope of
detailed physics (e.g., underlying correlations between the line flux
uncertainties, \citealt{loch2013}) that are accessible only by the theoretical method, it might provide a relevant benchmark for the latter. In this paper, we 
explore the uncertainty assessment based on observed spectra.

This is the third part of a series of papers centered on the Fe-L shell modeling for X-ray astrophysics. In the first
\citep[][paper I]{gu2019} and the second \citep[][paper II]{gu2020} parts, we presented respectively a theoretical model of the Fe-L complex 
spectrum for collisional plasma and an experimental benchmark of several key transitions. In paper II, we further carried out
a comprehensive analysis of the 600~ks Chandra High Energy Transmission Grating (HETG) of the Capella corona with 
a peak temperature of $0.5-0.6$~keV using the advanced
model. The Capella data features a photon-rich ($\sim 1.1 \times 10^{6}$ counts), line-rich ($>750$ lines), and well-resolved ($\sim 1.2$~eV resolution at 800~eV and $\sim 34.7$~eV at 6000~eV) spectrum. The
instrument calibration and astrophysical modeling (see \S~\ref{sec:method}) are reasonably well understood. These make the 
Capella data one of the 
best candidates for the 
study of model uncertainties. In order to cover the high temperature range, we also include in the test 
the 110~ks HETG data of 
HR~1099 corona and the 289~ks {\it Hitomi} data of the Perseus cluster. HR~1099 is a RS CVn binary with a broad 
coronal temperature
distribution within $1-3$~keV \citep{huenemoerder2013}. The Perseus cluster of galaxies is the brightest cluster
in the X-ray sky, the main source is the diffuse intracluster medium with a peak temperature of 4~keV \citep{hitominature}.

This paper is organized as follows. Section~\ref{sec:method} presents the new approach to assess the uncertainties in line fluxes
of a collisional spectral model based on observed data. We attempt to decouple it from the uncertainties in the ionization 
concentration which constitutes another major error component in the model. Section~\ref{sec:application} applies the obtained uncertainties to the science interpretation of spectra to be obtained with future XRISM and Athena missions \citep{Guainazzi2020}.
Throughout the paper,
the errors are given at a 68\% confidence level.

\begin{figure*}[!htbp]
\centering
\resizebox{1.0\hsize}{!}{\includegraphics[angle=0]{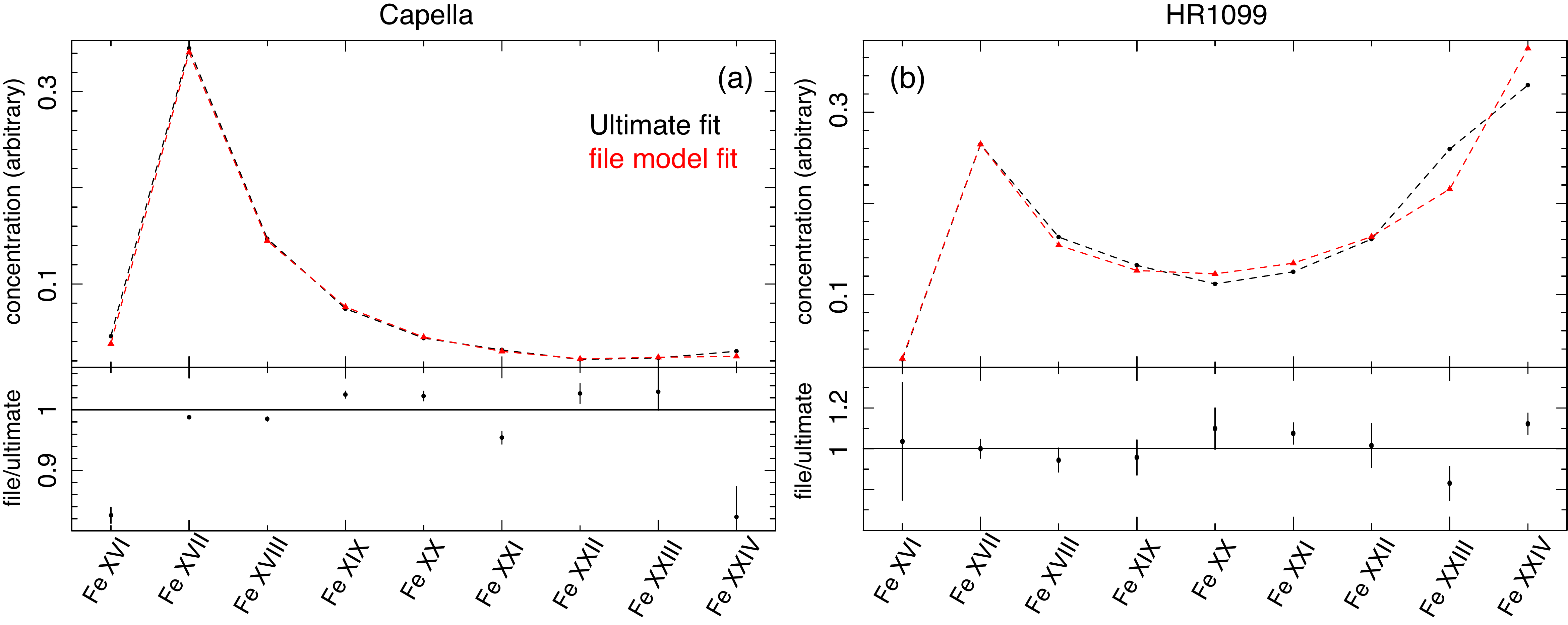}}
\caption{ The charge state distributions of Fe ions calculated from the multi-temperature models of Capella (a) and HR~1099 (b). The black and red
dashed lines are obtained from the fits with the best physical model (i.e., model 1 in \S~\ref{sec:model}) and with the file model which best describes the observation. The lower panels show the ratios
between the two models. }
\label{fig:ibal}
\end{figure*}

\section{Method and results}
\label{sec:method}

In Paper II, we carried out a global, self-consistent modeling and detailed fit to the {\it Chandra} HETG data of 
Capella, based on the SPEX code \citep{kaastra1996} with improvements on atomic database 
made in Papers I and II. The collisional excitation and dielectronic recombination cross sections of the Fe L-shell
species are updated to the modern R-matrix calculations (\ion{Fe}{XVII} from \citealt{liang2010}, \ion{Fe}{XVIII}
from \citealt{witt2006}, \ion{Fe}{XIX} from \citealt{butler2008}, \ion{Fe}{XX} from \citealt{witt2007}, \ion{Fe}{XXI} from 
\citealt{badnell2001}, \ion{Fe}{XXII} from \citealt{liang2012}, \ion{Fe}{XXIII} from \citealt{fern2014}, and \ion{Fe}{XXIV}
from \citealt{liang2011}), augmented by a large scale distorted wave calculation 
when R-matrix results are not available. The measured cross sections of \ion{Fe}{XVI} and \ion{Fe}{XVII} excitation
and dielectronic recombination from an Electron Beam Ion Trap experiment \citep{shah2019} have also been incorporated. 
These improvements are included in SPEX version 3.06.
Results on the Capella spectrum were shown for the
fits with three kinds of models ordered by the level of sophistication: the `baseline', the `advanced', and the `ultimate' 
models. 
The `baseline' model utilizes a combination of 18 collisional ionization equilibrium (CIE) components, characterized by the common elemental abundances, to approximate 
the multi-temperature structure of the coronal plasma. The model further takes into account the effect of
interstellar absorption, resonant scattering, and astrophysical turbulence, as well as various systematic uncertainties
from the instrumental calibration on, e.g., the effective area, energy scale, and line spread function. The 
`advanced' model improves further from the `baseline', by allowing the elemental abundances of different CIE components to vary
freely, decoupling the temperatures used for ion concentration calculation from the ones for spectral evaluation, and setting
the plasma density as a free parameter. The `ultimate' model is basically the same as the `advanced' model, except that the former
further applies a fix to possible wavelength errors in the code by comparing the data in SPEX with those in APEC and Chianti.
The fit with the ultimate model becomes the best of the three, however, the final C-statistic value (40281 for expected value 
of 7137) remains formally unacceptable, revealing remaining 
uncertainties in the best available atomic data and plasma codes.

The differential emission measure distribution obtained with the ultimate model shows a peak at $\sim 0.6$~keV (see
Figure~11 in paper II), 
which in general agrees with the previous measurements using the {\it Chandra} HETG and {\it XMM-Newton} Reflection Grating Spectrometer (RGS) data (e.g., \citealt{gu2006}). The chemical abundances of C, O, Ne, and Ni are found to be sub-solar,
while the abundances of N, Na, Mg, Al, and Cr are solar or above (see Table~4 in paper II). These results agree within 
the uncertainties with the values reported in \citet{gu2006}, except for the Ni abundance which is 40\% higher in their work.
The Fe, Si, S, Ar, and Ca abundances are set free to vary among different CIE components (Figure~12 in paper II), so that they 
cannot be compared directly with the previous reports in which these abundances were tied among all the components.  
The electron density of the stellar corona is determined to be $<1.4\times 10^9$ cm$^{-3}$, which is in good agreement
with the previous results, e.g., $< 2.4\times 10^9$ cm$^{-3}$ by \citet{ness2001} and $< 7\times 10^9$ cm$^{-3}$ by 
\citet{mewe2001} obtained with the {\it Chandra} Low Energy Transmission Grating Spectrometer (LETGS) data.

%The `ultimate' model fully took into account various systematic uncertainties 
%from the instrumental calibration, e.g., the effective area, energy scale, and line spread function, as well as
%astrophysical effects such as the multiple temperature, astrophysical turbulence, absorption/scattering, density diagnostics,
%and possible systematic shift from ionization balance. Though the ultimate model improves significantly from the other two,
%the final C-statistic value (40281 for expected value of 7137) remains formally unacceptable, revealing remaining 
%uncertainties in the best available atomic data and plasma codes.

Here we introduce a new method to assess the atomic uncertainty on line emissivities using the data from real observations. 
First, we revise the `ultimate' spectral model obtained in Paper II, incorporating additional degrees of freedom that
could set the ion concentration and the emissivities of strong lines as free parameters in the fits (see \S~\ref{sec:model} for details). 
By fitting the revised model to the observed high-quality spectra, we can anchor the obtained deviations between theoretical
values and the actual data to the underlying uncertainties in the atomic modeling. 

%First, 
%we revise the best present spectral model obtained in Paper II, by incorporating additional degrees of freedom in the 
%model fine-tuning the ionization balance 
%and the line emissivities. Then, by fitting the real data for a large sample of lines, we can anchor the obtained deviations 
%from the theoretical values to the underlying uncertainties in atomic physics.

This work focuses on a set of ions of interest (IOIs), which are \ion{Fe}{XVI} $-$ \ion{Fe}{XXIV}, \ion{Ni}{XIX}, \ion{N}{VII},
and He- and H- like O, Ne, Na, Mg, Al, Si, and S. Though the Ar and Ca lines are also visible in the Capella spectrum, 
they are not included in the IOIs as the quality of {\it Chandra} grating spectrum at these lines are not sufficient for a
robust study.

\begin{figure*}[!htbp]
\centering
\resizebox{1.0\hsize}{!}{\includegraphics[angle=0]{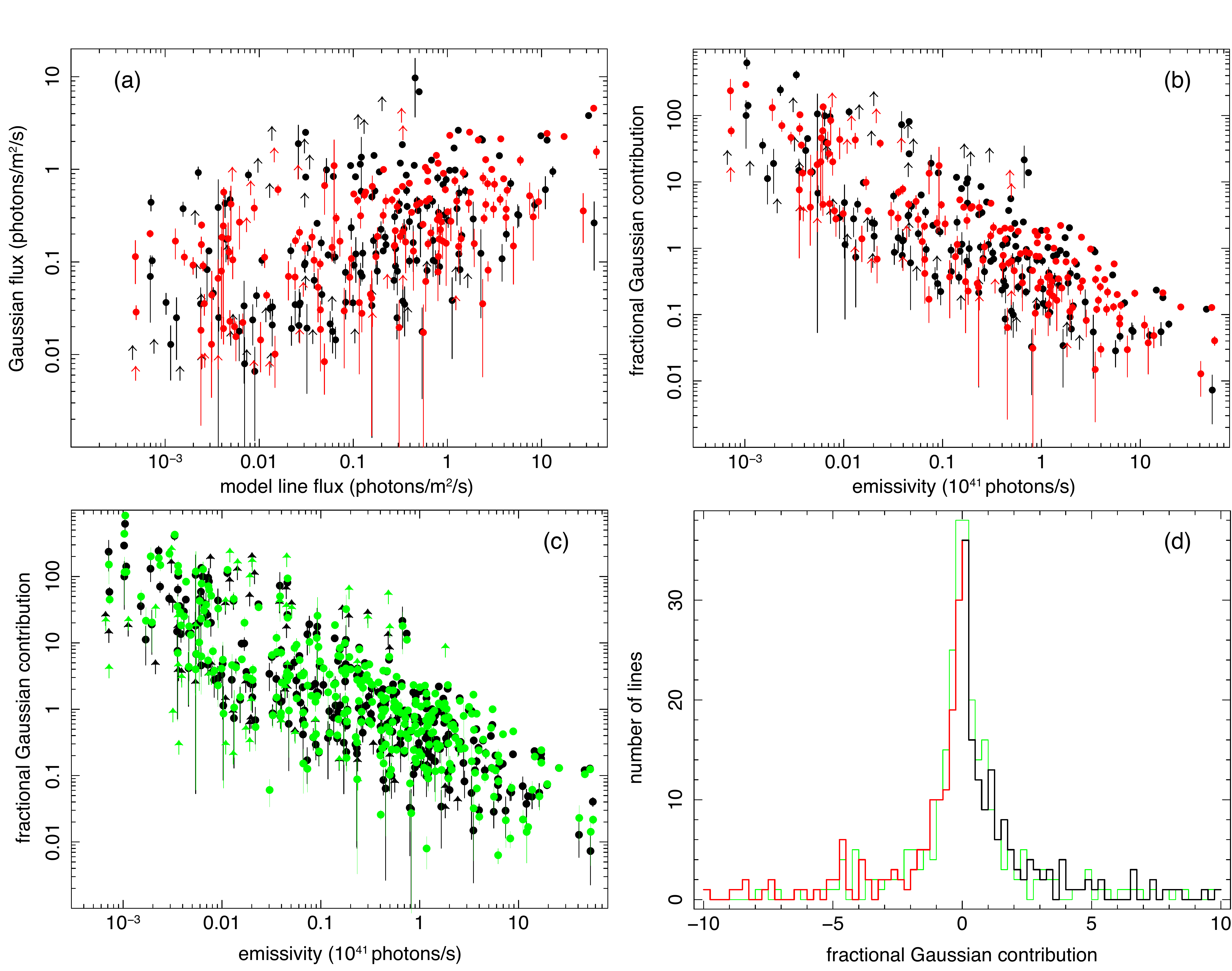}}
\caption{Line flux uncertainties in the modeling of the Capella HETG spectrum. (a) Absolute fluxes of the additional Gaussian components 
(or the differential fluxes between model 2 and model 1, see \S~\ref{sec:model} for details) plotted against the theoretical fluxes of the corresponding emission lines. Each point represents one line. The black and red data points are the Gaussian components with positive and negative normalizations, respectively. Arrows show the upper limits. (b) Fractional contributions of the Gaussian components to the total fluxes shown as a function of line emissivities. (c) Same as (b), but shows a comparison between the results with line widths fixed (black) and line widths free to fit (green). The black data points include both the black and red data points in (b). (d) Histogram of the Gaussian contribution. The black and red data are the results with positive and negative Gaussian 
normalizations from the fit with fixed line widths, and the green data are the ones with the free line widths.}
\label{fig:ratio_capella}
\end{figure*}

\begin{figure}[!htbp]
\centering
\resizebox{1.0\hsize}{!}{\includegraphics[angle=0]{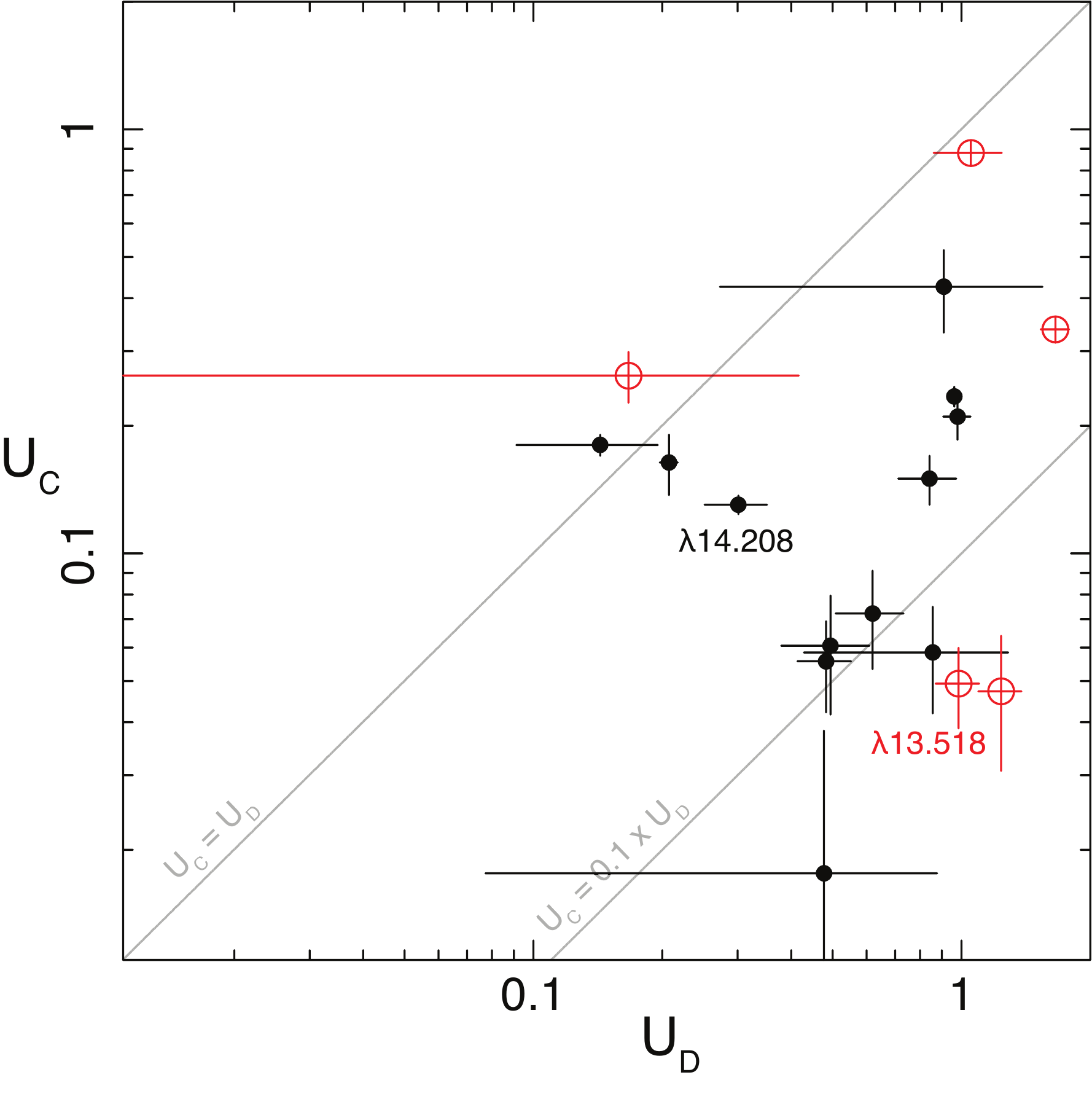}}
\caption{Fractional uncertainties of \ion{Fe}{XVIII} and \ion{Fe}{XIX} lines obtained in \citet[U$_{\rm D}$]{desai2005} with the Capella spectrum compared
with those derived in the current work (U$_{\rm C}$). Black and red data points represent \ion{Fe}{XVIII} and \ion{Fe}{XIX} lines, respectively. }
\label{fig:desai}
\end{figure}

\begin{figure*}[!htbp]
\centering
\resizebox{1.0\hsize}{!}{\includegraphics[angle=0]{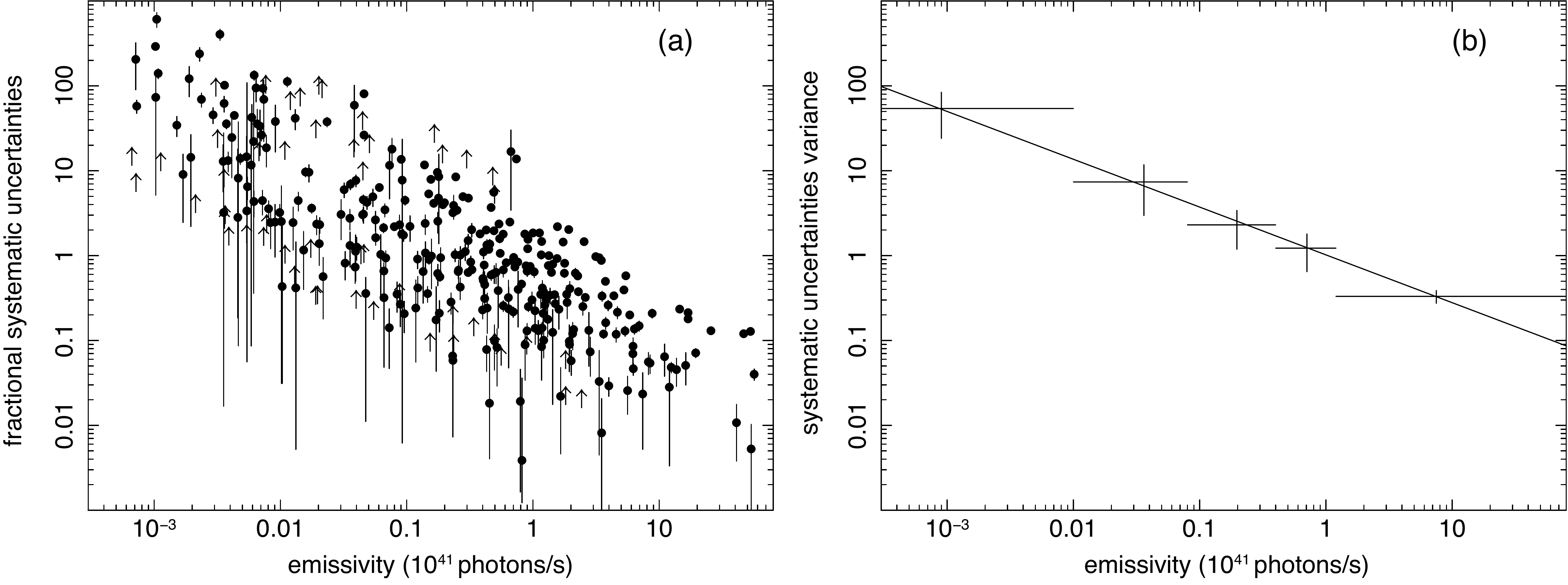}}
\caption{(a) Systematic uncertainties of the line flux modeling of the Capella spectrum, derived by subtracting the 
statistical uncertainties from the total uncertainties obtained above (Figure~\ref{fig:ratio_capella}). 
 (b) Variances of the systematic uncertainties as a function of 
the theoretical line emissivities. The solid line shows the analytic fit with Eq.~\ref{eq:rel}.
}
\label{fig:ratio_fwhm}
\end{figure*}

\begin{figure*}[!htbp]
\centering
\resizebox{1.0\hsize}{!}{\includegraphics[angle=0]{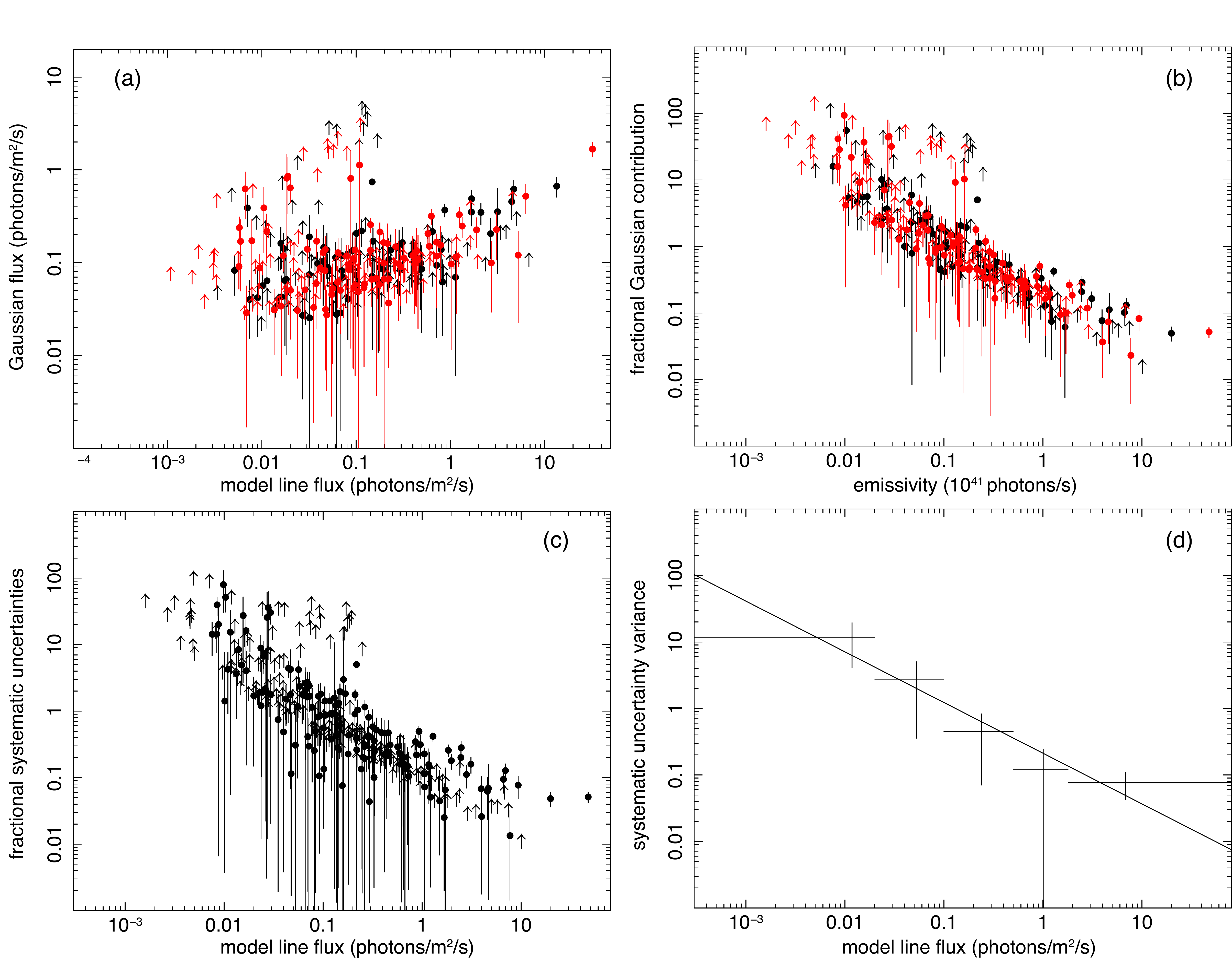}}
\caption{(a \& b) Same as Fig.~\ref{fig:ratio_capella} panels a and b but for HR~1099. (c \& d) Same as
Fig.~\ref{fig:ratio_fwhm} panels a and b but for HR~1099.}
\label{fig:ratio_hr1099}
\end{figure*}

\begin{figure*}[!htbp]
\centering
\resizebox{1.0\hsize}{!}{\includegraphics[angle=0]{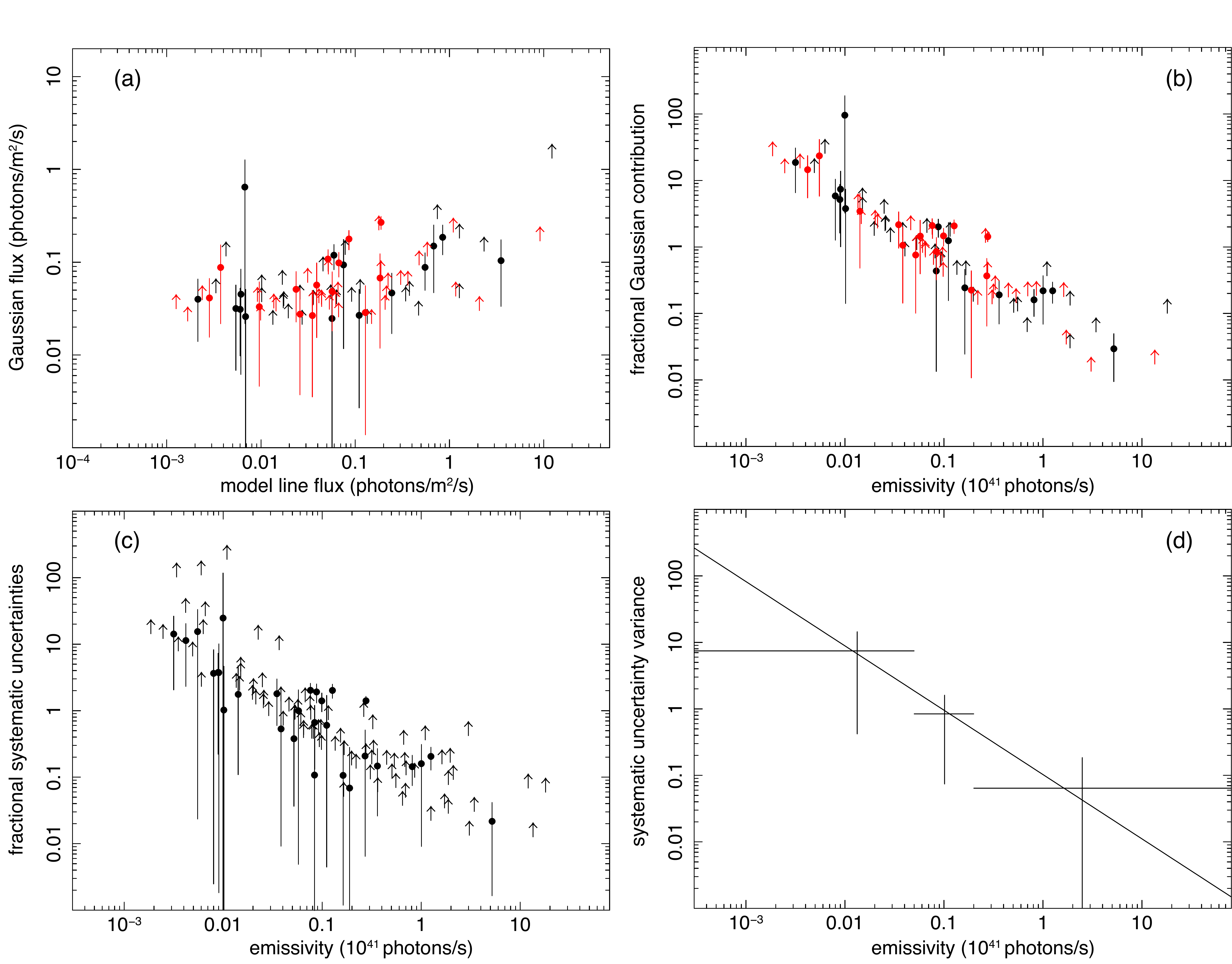}}
\caption{(a \& b) Same as Fig.~\ref{fig:ratio_capella} panels a and b but for the Perseus cluster with the Hitomi data.
(c \& d) Same as Fig.~\ref{fig:ratio_fwhm} panels a and b but for the Perseus cluster.}
\label{fig:ratio_perseus}
\end{figure*}

\begin{figure*}[!htbp]
\centering
\resizebox{1.0\hsize}{!}{\includegraphics[angle=0]{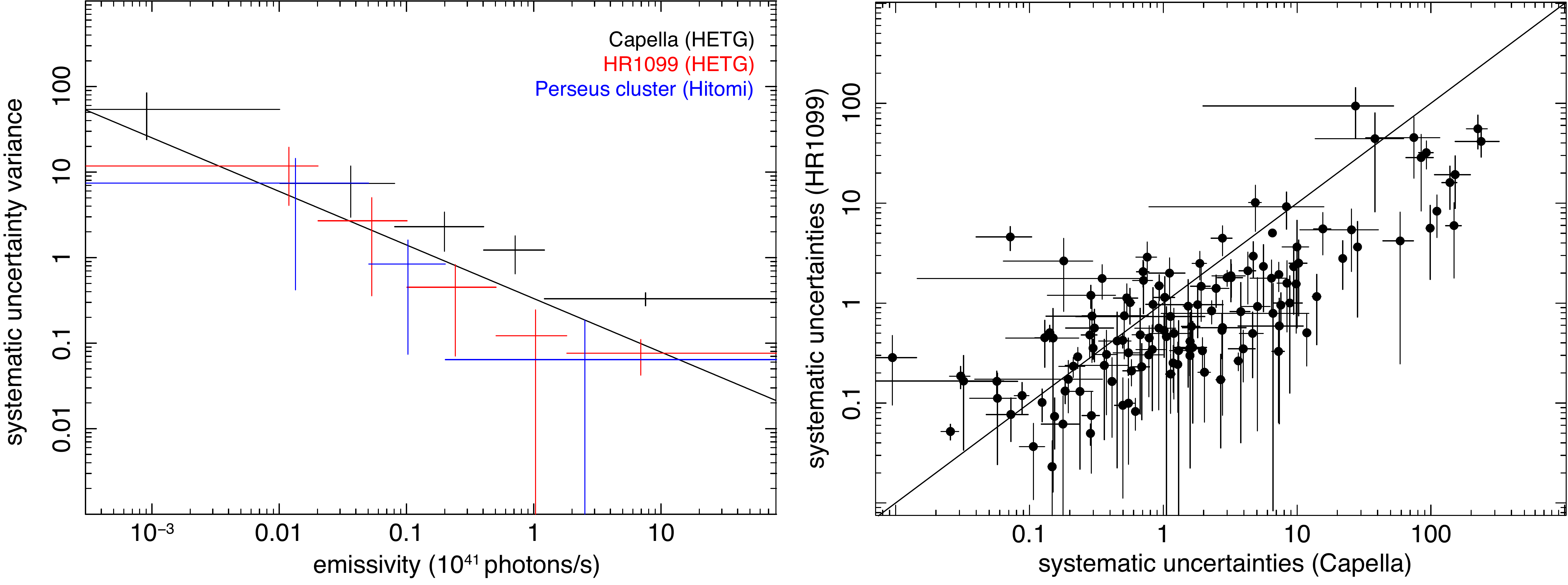}}
\caption{(left) Variances of the systematic uncertainties in the line intensities for Capella (black), HR~1099 (red),
and the Perseus cluster (blue). The solid line shows the analytic fit to the combined data (see Table~\ref{tab:fitpara}). 
(right) Systematic uncertainties obtained with the Capella spectrum plotted against those 
obtained with
the HR~1099 data for the same set of transitions. }
\label{fig:ratio_all}
\end{figure*}

\begin{figure}[!htbp]
\centering
\resizebox{1.0\hsize}{!}{\includegraphics[angle=0]{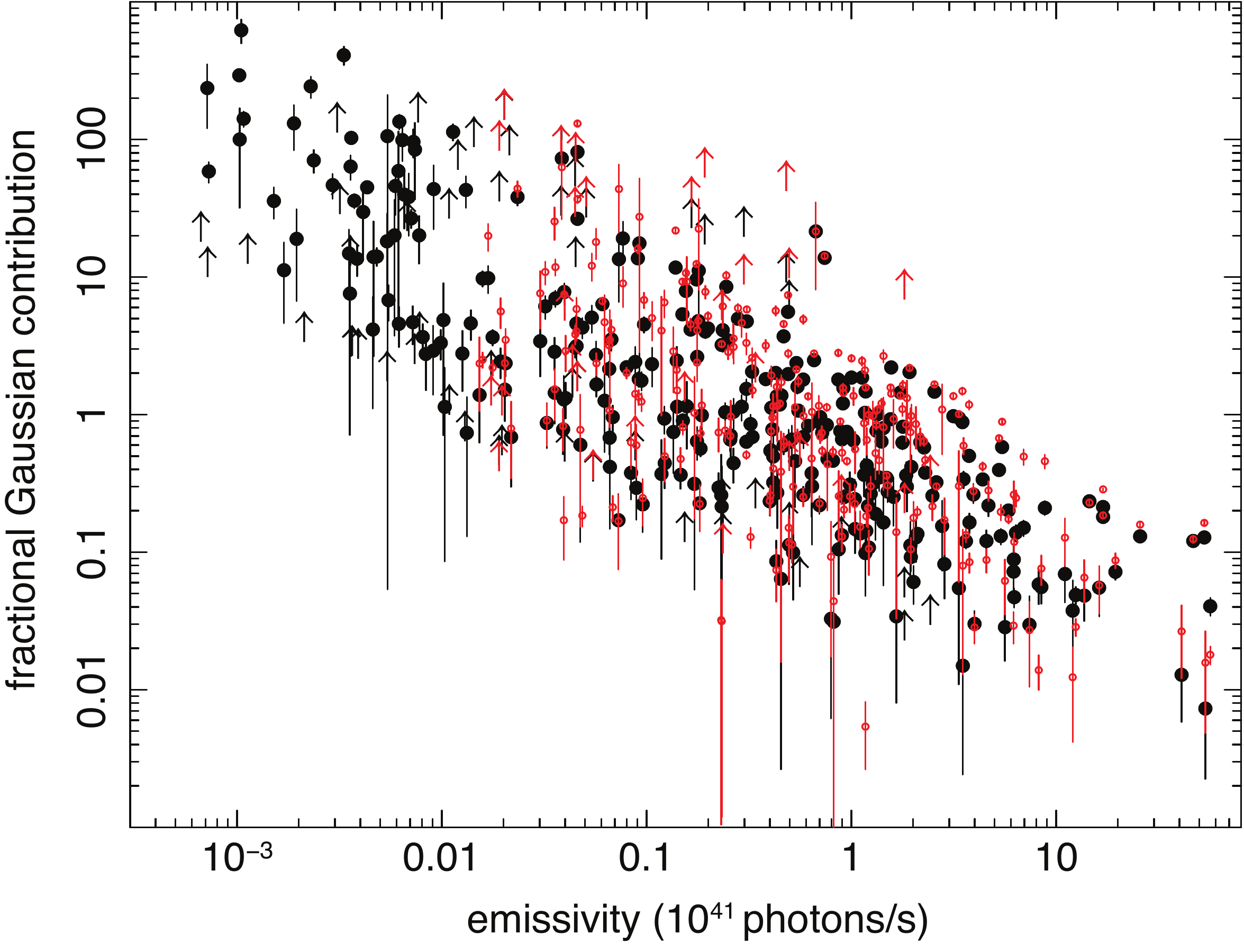}}
\caption{The fractional line flux uncertainties of Capella with the original model (Fig.~\ref{fig:ratio_capella}, black)
compared with the uncertainties obtained with the model excluding the weak transitions (red). See text for details.}
\label{fig:ignoreweak}
\end{figure}

\begin{figure*}[!htbp]
\centering
\resizebox{1.0\hsize}{!}{\includegraphics[angle=0]{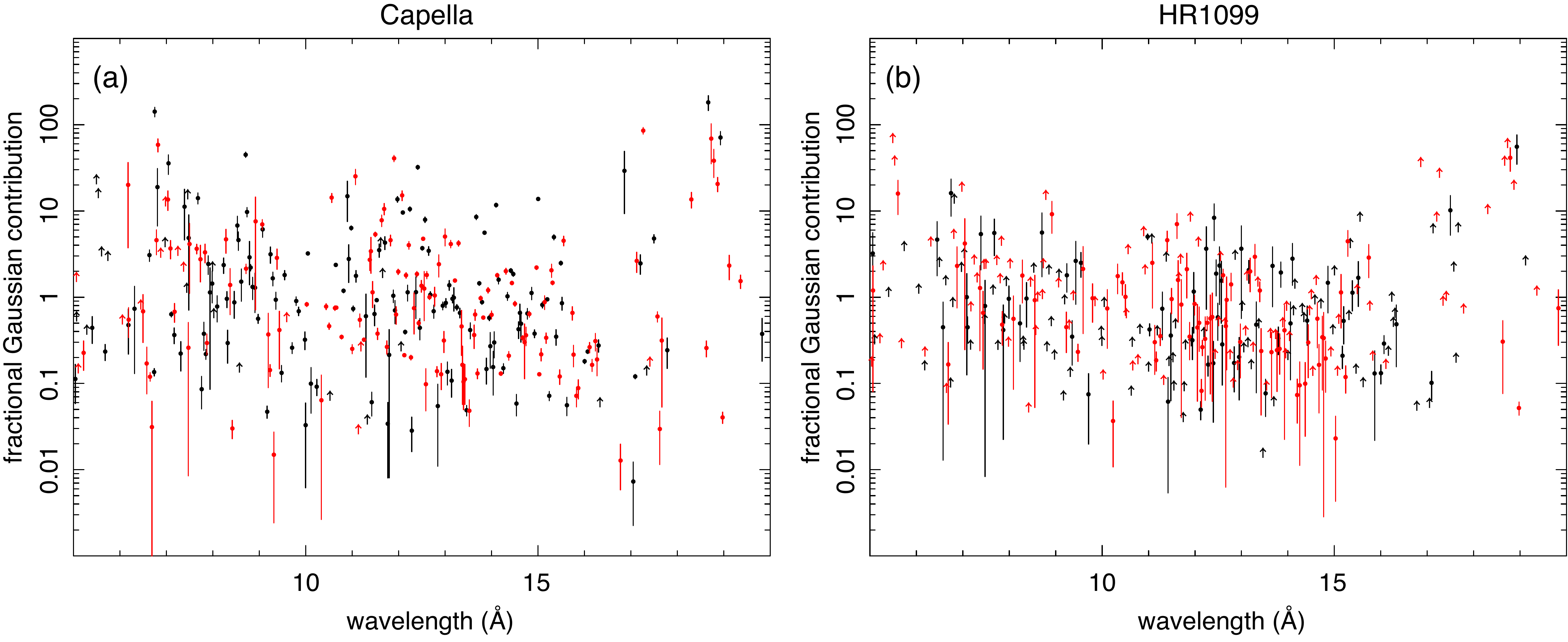}}
\caption{Fractional Gaussian contributions to the total line fluxes plotted as a function of line wavelengths for the Capella (a) and HR~1099 (b)
data. The black and red data points are the Gaussian components with positive and negative normalizations,
respectively.}
\label{fig:wavelength}
\end{figure*}

\begin{figure*}[!htbp]
\centering
\resizebox{1.0\hsize}{!}{\includegraphics[angle=0]{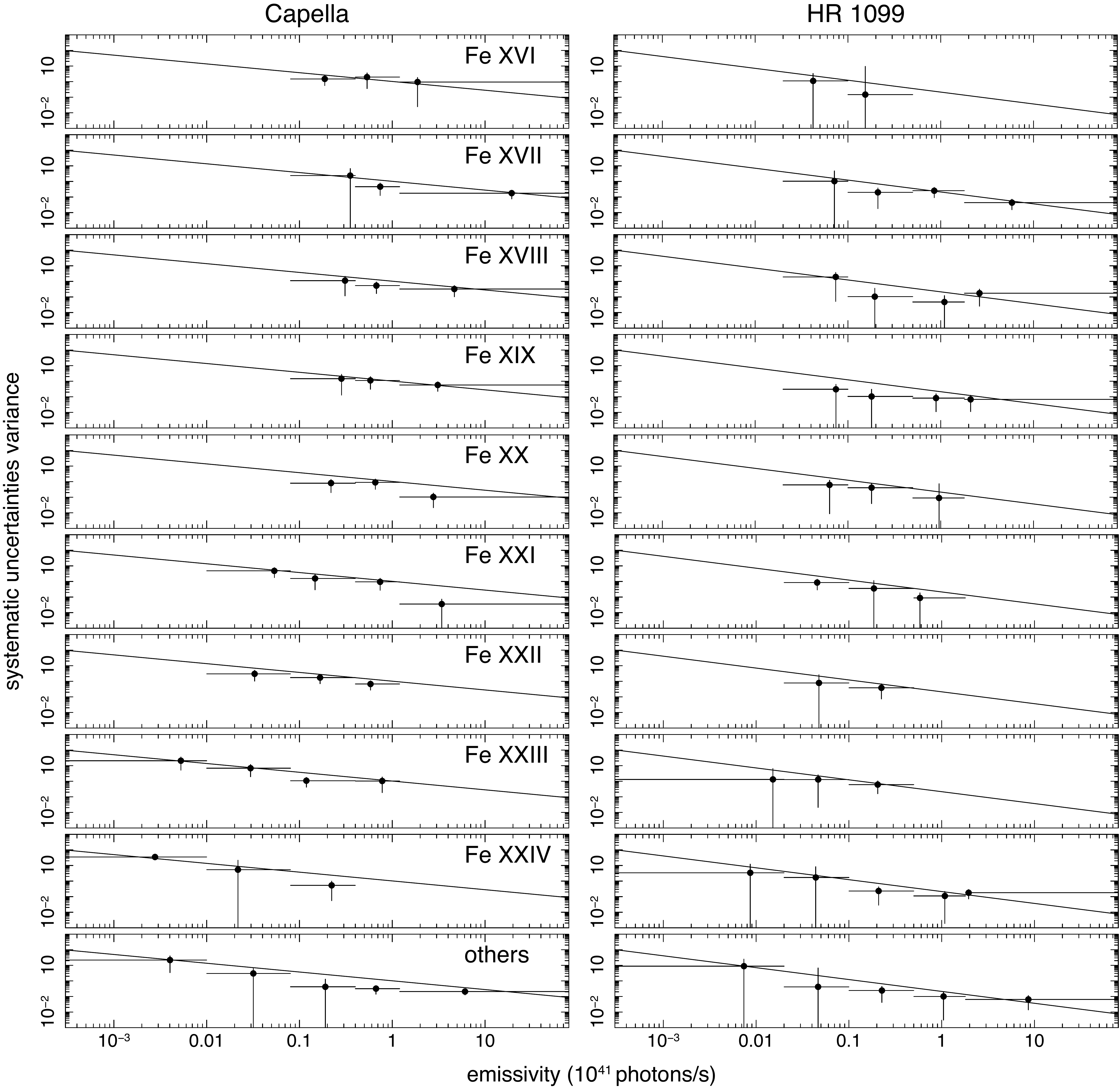}}
\caption{Variances of systematic uncertainties plotted as a function of line emissivities for individual ions based on
the Capella (left) and HR~1099 (right) spectra. The solid lines are the same as in Figs.~\ref{fig:ratio_fwhm} (b) and 
\ref{fig:ratio_hr1099} (d).}
\label{fig:ratio_ion}
\end{figure*}

\begin{figure*}[!htbp]
\centering
\resizebox{1.0\hsize}{!}{\includegraphics[angle=0]{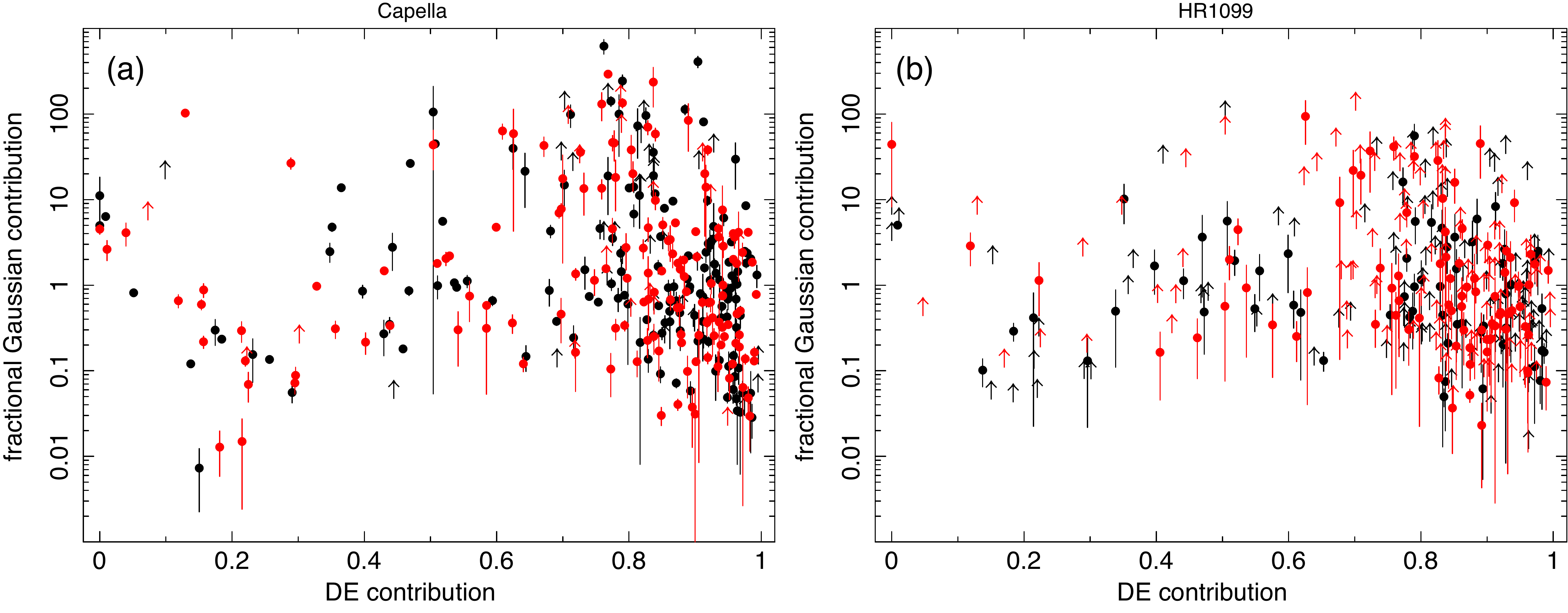}}
\caption{Fractional Gaussian contributions to the total line fluxes plotted as a function of contributions to the upper level formation
by direct collisional excitation for Capella (a) and HR~1099 (b). The black and red data points are the Gaussian components with positive and negative normalizations, respectively. }
\label{fig:ratio_de}
\end{figure*}

\begin{figure*}[!htbp]
\centering
\resizebox{1.0\hsize}{!}{\includegraphics[angle=0]{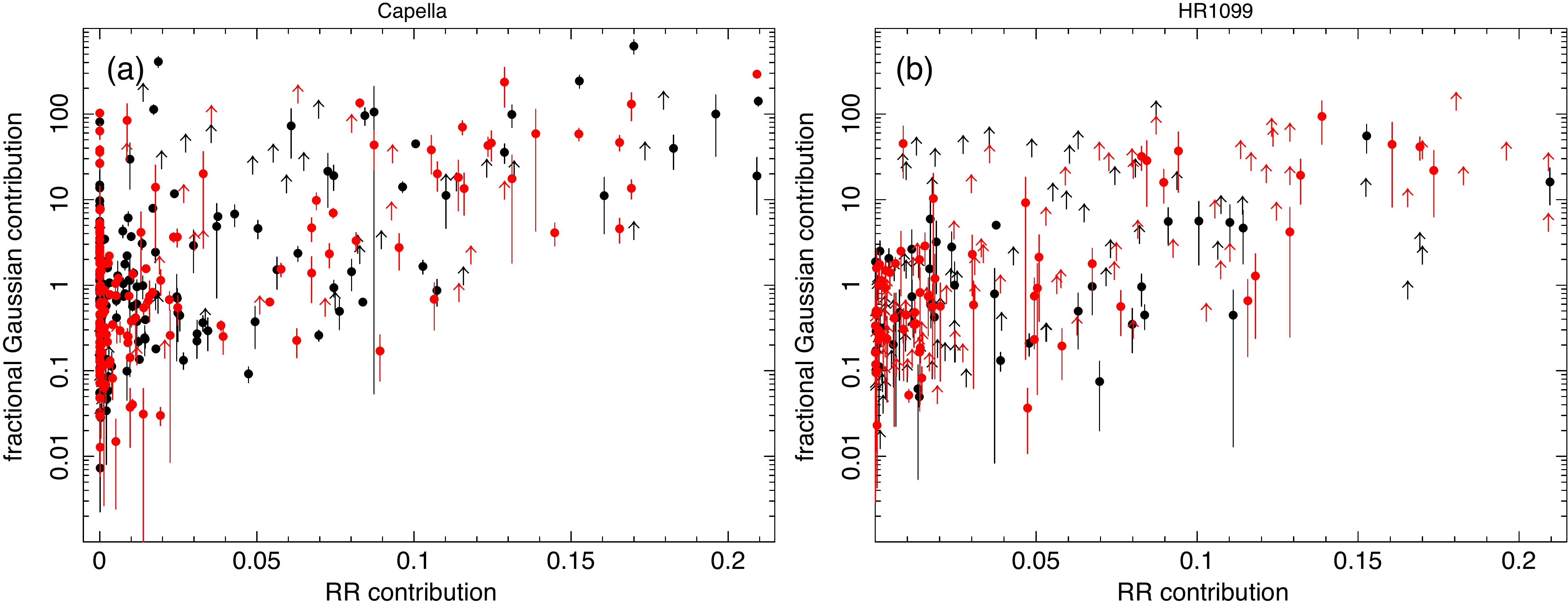}}
\caption{Same as Fig.~\ref{fig:ratio_de} but for radiative recombination.}
\label{fig:ratio_rr}
\end{figure*}

\begin{figure*}[!htbp]
\centering
\resizebox{1.0\hsize}{!}{\includegraphics[angle=0]{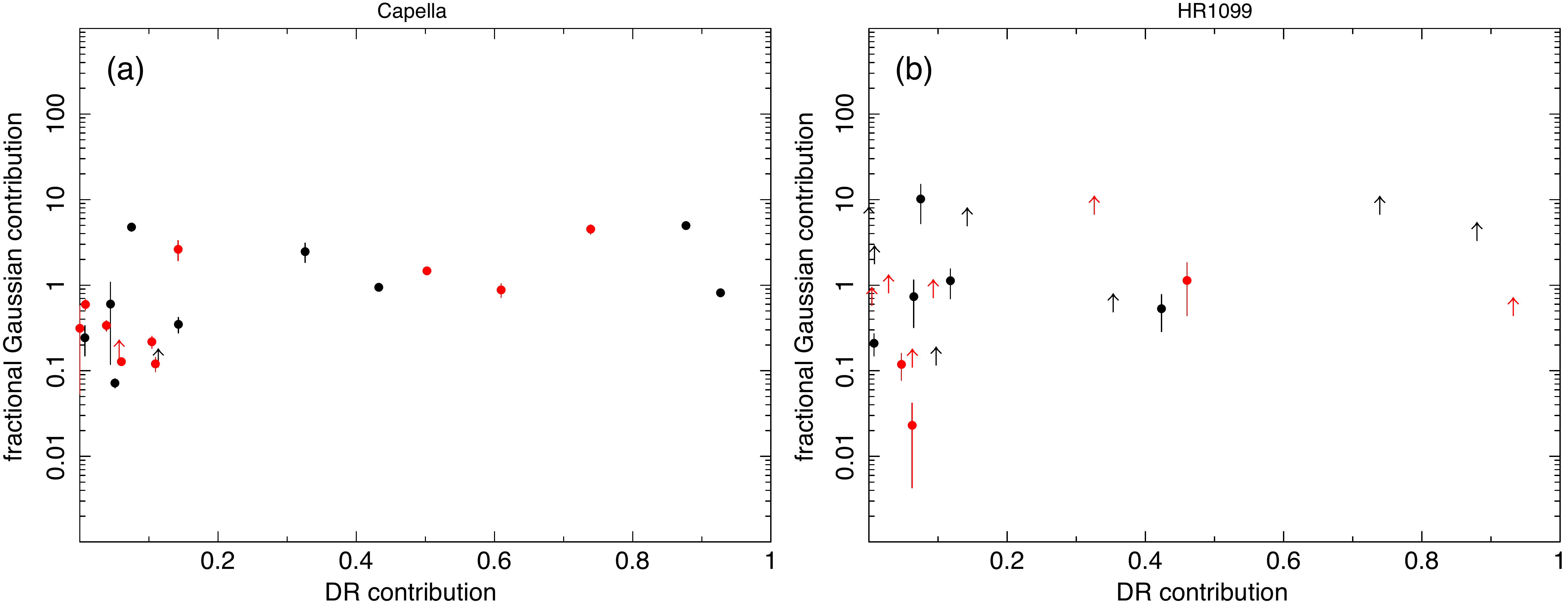}}
\caption{Same as Fig.~\ref{fig:ratio_de} but for dielectronic recombination.}
\label{fig:ratio_dr}
\end{figure*}

\begin{figure*}[!htbp]
\centering
\resizebox{1.0\hsize}{!}{\includegraphics[angle=0]{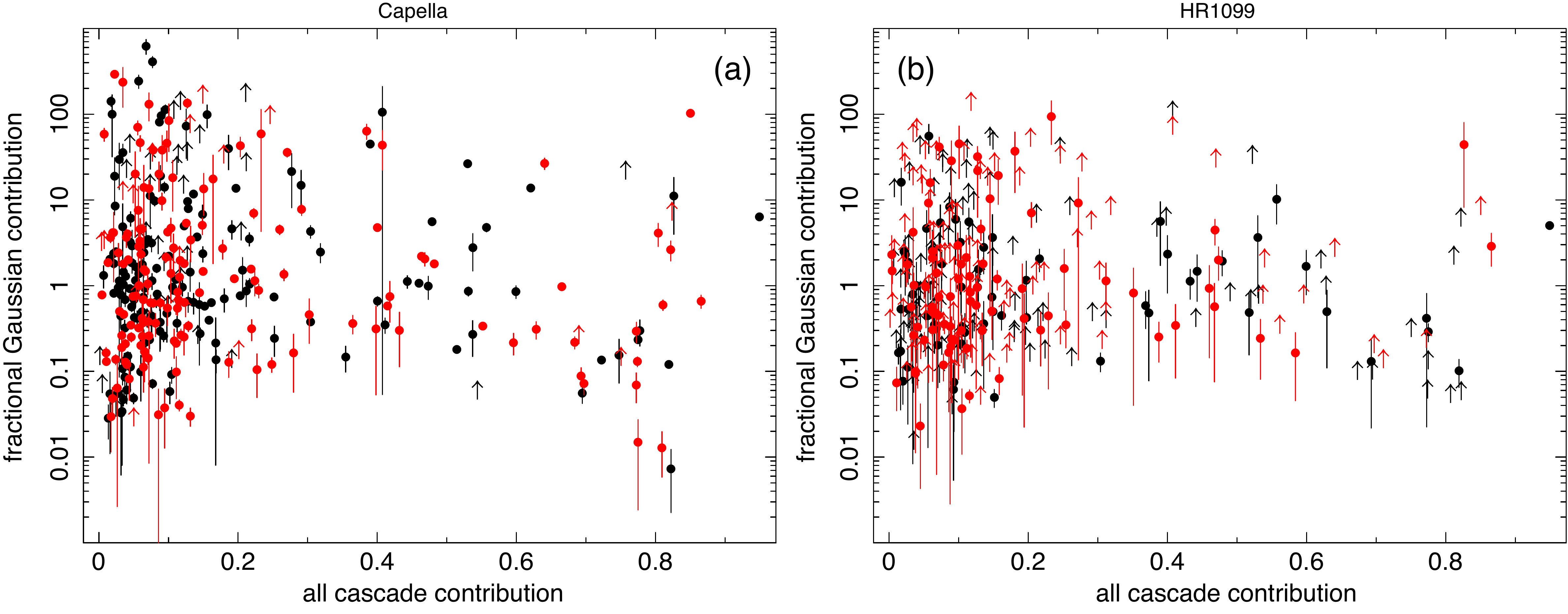}}
\caption{Same as Fig.~\ref{fig:ratio_de} but for radiative cascade from higher levels.}
\label{fig:ratio_allcas}
\end{figure*}

\subsection{Data}
\label{sec:data}
This work is based on high quality spectroscopic observations of Capella, HR~1099, and the Perseus cluster, observed 
with the {\it Chandra} HETG and the {\it Hitomi} Soft X-ray Spectrometer (SXS). 
Capella and HR~1099 are both representative bright stellar corona objects: the former has a peak temperature of $0.5-0.6$~keV
\citep{p2001, behar2001, desai2005, gu2006} and the latter is substantially hotter ($\sim 1-3$~keV, \citealt{ayres2001, huenemoerder2013}). Capella is the best target for testing models of 
\ion{Fe}{XVI} $-$ \ion{Fe}{XX}, while HR~1099 is appropriate for higher ionization states, e.g., \ion{Fe}{XX} $-$ \ion{Fe}{XXIV}. 
The Medium Energy Grating (MEG) spectra in the wavelength range of $3-32$~{\AA} are used for the two stellar objects.
MEG data have an energy resolution of 1.2~eV at 800~eV, where the Capella spectrum peaks.  
The High Energy Grating (HEG) is more suited for sources of higher energies, since it cannot cover energies 
below 800~eV. Due to the instrumental efficiency, the 
HEG count rates of the two stellar objects in $5-20$~{\AA} are 
lower, by a factor of 3.8 (Capella) and 2.8 (HR~1099), than those of the MEG data. To prevent systematic uncertainties
by cross calibration between MEG and HEG, we only use the MEG data for the two coronal objects.
Similarly, we do not 
include the data from Low Energy Transmission Grating (LETG) in this work. Although LETG provides the best spectral resolution
for soft X-ray at $> 50$~{\AA}, its resolution gradually gets worse at shorter wavelengths. At 800~eV the LETG resolution 
becomes 2.8~eV. Therefore the MEG has the best combination of spectral resolution and data statistics for the source spectra. 
The raw data were reduced using the CIAO v4.10 and calibration database (CALDB) v4.8. The $chandra\_repro$ script is used for the data screening and production of spectral files. The multiple Capella spectra and
the associated response files are combined using the CIAO $combine\_grating\_spectra$ tool. The spectral analysis is carried 
out with SPEX version 3.06 \citep{kaastra1996, kaastra2020}.

Our work relies on the critical assumption that most of the uncertainties in instrumental calibration can be 
properly dealt with in the analysis. Previous reports show that the systematic uncertainties are about 8\% in the HETG effective area 
calibration and $10^{-5}$ in the wavelength scale 
\footnote{\url{https://cxc.harvard.edu/cal/summary/Calibration_Status_Report.html}}. 
To overcome the possible deviations in the calibration, we incorporate specific functions as follows in the modeling of the 
grating spectra. 
The possible residual calibration errors in the MEG effective area for continuum are corrected by 
multiplying the main spectral components by the SPEX {\it knak} model, 
which defines
a set of piecewise power-law correction functions with grid points of 1, 3, 6, 10, 14, 18, 26, 32, and 38~{\AA}. We also incorporate
a neutral oxygen absorption model to model the instrumental uncertainty at the \ion{O}{I} edge ($22.6 - 22.9$~{\AA}). 
By making several iterations between a fit with 100 eV-wide bins and a fit with the optimal binning, the best-fit {\it knak} 
and \ion{O}{I} edge models are determined.
The systematic uncertainties in the wavelength scale have been corrected by applying a redshift component on the astrophysical 
model. The MEG 
line spread function is modeled with the arbitrary line broadening model {\it vpro}, with a
profile shape calculated from the observed \ion{O}{VIII} Ly$\alpha$ line at $\sim 19$~{\AA}. A Wiener filter has been
applied to the background dataset obtained with the standard pipeline. This filter minimizes the background noise by using
a Fourier transform.

Above instrumental modeling and fine-tuning are introduced in order to provide
a platform that allows a fair comparison between two different sets of atomic line modeling. The priority is therefore 
to correct the possible energy-dependent calibration residuals and biases throughout the wavelength range, 
rather than to achieve the absolute calibration precision that is otherwise needed for the measurements of astrophysical parameters. 
For instance, 
although the {\it knak}
component might not be able to provide the absolute correct value for the instrumental effective area, it is sufficient to remove 
the wavelength-dependent biases in the continuum modeling.

%We also smooth out the noisy features
%in the default background dataset with a Wiener filter, which determines the noise level by a Fourier transform.

A similar method has been applied to the {\it Hitomi} observations of the Perseus cluster, which has a peak temperature of $\sim 4$~keV.
The micro-calorimeter data has a resolution of $\sim 5$~eV in the energy range of $2-10$~keV ($1.2-6.2$~{\AA}). The data
screening and calibration corrections are identical to those reported in \citet{hitomiatomic}. The {\it Hitomi} data is used as an essential crosscheck 
at shorter wavelengths for our analysis with the stellar sources.

\subsection{Spectral modeling}
\label{sec:model}
We utilize the `ultimate' model described at the beginning of \S~\ref{sec:method} as a template (hereafter model 1) for 
the Capella and HR~1099 spectra. The differential emission measure of the quiescent coronal plasma
can be well approximated by the sum of multiple CIE components (e.g., \citealt{brickhouse2000, desai2005}). Model 1 
contains 18 independent CIE components forming a quasi-continuous emission measure distribution. The temperatures of the 
18 components
are set by Eq.~5 of Paper II to the fixed values of 0.12~keV, 0.15~keV, 0.18~keV, 0.21~keV, 0.25~keV, 0.30~keV, 0.36~keV, 0.43~keV, 0.51~keV, 0.61~keV, 0.75~keV, 0.92~keV, 1.15~keV, 1.47~keV, 1.94~keV, 2.70~keV, 4.07~keV, and 7.00~keV. 
The emission measure of each component is 
set free to vary in the fits.  %It is a collection of 18 
%collisional ionization equilibrium (CIE) components, with consecutive temperatures from $0.1-10$~keV (see Eq.~5 of paper II), 
%for approximating the coronal emission measure distribution. 
The Fe abundance of each component can also vary independently, 
and the Si, S, Ar, and Ca are fit quasi-independently by properly grouping of the temperature components. 
As shown in Figures~11 and 12 of Paper II, both the emission measure and abundances of each temperature component 
can be well constrained from the fit, because the former can be derived from the total line plus continuum emissivity, 
while the latter is determined mostly from several relevant lines.
Our model further takes into account a set 
of astrophysical effects, including the density-sensitive lines, turbulent broadening at each temperature, 
systematic line shift, 
systematic bias from ionization equilibrium, the neutral and ionized interstellar absorption, and resonant scattering. 

The model applied to the {\it Hitomi} spectrum of the Perseus cluster is composed of three independent temperature components
\citep{hitomiatomic}, each with temperature, emission measure, turbulence velocity, and Fe abundance free to fit. Our model
takes into account effects of the AGN emission, resonance scattering, charge exchange, and the Galactic absorption.

An accurate continuum modeling is essential for assessing line uncertainties. The continuum of model 1 is composed by 
the Bremsstrahlung, two photon emission, and free-bound radiation components with a quasi-continuous electron temperature distribution described above. As shown in Figure~B.1$-$3 of Paper II, the model continuum agrees within $\sim 10$\% with the Capella spectrum in $4-18$~{\AA}. The continuum at $>18$~{\AA} seems to be overestimated by $20$\% on average, but this affects only $< 10$\% of the total lines since most L-shell transitions are in the range of $8-18$~{\AA}. As shown in Figure~23 of \citet{hitomiatomic}, the continuum model of the Perseus cluster is in agreement within $5-10$\% with the data in the entire band of the observed data. The non-thermal component is found negligible in Capella and the Perseus cluster. The possible small residuals due to instrumental effective area calibration issues have been corrected with the {\it knak} component described in \S~\ref{sec:data}. 

%Based on a manual
%check, this model also contains corrections of a set of wavelength errors in the atomic code (see Table~B.2 of paper II), which
%are found essential at least for the Capella analysis. 

\subsubsection{Uncertainties on ion concentration}

Before examining the uncertainties in the model line fluxes, we would like to address the possible deviations in the 
ion concentration (or charge state distribution) calculation. The uncertainties in ion concentration (with respect to the 
present equilibrium values, e.g., 
\citealt{u2017}) are in general expected to be another major component in the total error budget \citep[e.g.,][]{foster2020}.
To estimate uncertainties in ion concentration, first
we need to verify the assumption of collisional ionization equilibrium for the sources. This could be done by setting 
{\it rt} of the SPEX model ($\it rt$ = 1 for an equilibrium case) a free parameter in the fit, in such a way the temperature used for calculating the ionization 
balance is decoupled from the temperature for the evaluation of rate coefficients (e.g., the excitation rates). For 
Capella and HR~1099, {\it rt} is determined to be $0.99 \pm 0.01$ and $1.0 \pm 0.01$, and \citet{hitomiatomic} reported
a near unity value of $0.98 \pm 0.01$ for the Perseus cluster. This means that the non-equilibrium effects are negligible 
for the three test sources.

As shown in Figure~\ref{fig:ibal}, the total ion concentration with model 1
can be determined by combining the ion concentration of each component weighed by the emission measure derived from the fit. 
To compare the 
model concentration with the observed one, we re-fit the spectra by allowing the model component of each IOI free to vary.
This can be done by the SPEX {\it file} model, which reads a spectral model from an ASCII file. The emission of IOIs are now modeled by multiple {\it file} models,
each contains the line plus continuum of one IOI from all the spectral components. The remaining 
non-IOI spectra are converted to another
{\it file} model. The astrophysical and instrumental corrections are applied to all the {\it file} models. This would allow
us to determine the absolute ion concentration directly from the observation which does not depend on the temperatures of the CIE components.
Through a fit
of the Capella and HR~1099 spectra, we determine the normalizations of the {\it file} components, and thus the deviations 
on ion concentrations from the values with model 1. For the Fe IOIs shown in Figure~\ref{fig:ibal},  the measured 
deviation is less than 20\% at worst, with an average value of 8\% for Capella and 
10\% for HR~1099. Similar values can be obtained with the other IOIs. The current estimate is also in good agreement with
the uncertainties obtained in \citet{hitomiatomic} for the Perseus cluster.

It should be noted that the measured deviation might not come fully from the ion concentration uncertainty alone. 
For several IOIs, the thermal emission is dominated by the strong emission lines. A possible systematic 
bias in the emissivities of the strongest transitions could therefore cause potential deviation in the measured concentration of the 
relevant IOI. In addition, the emissivities of radiative and dielectronic recombination and innershell ionization transitions are determined by the concentration of the neighbour ions. The changes of their emissivities due to the 
changes in relative ion concentration are not accounted in the current approach. As both effects would introduce 
extra deviations to the IOI concentration measurement, we consider the deviation shown in Figure~\ref{fig:ibal}
as a crude approximation to the actual atomic uncertainty in ion concentration.

%It should be noted that the above deviation might be a composition of different origins. Some IOIs might be biased
%astrophysically, through possible abrupt thermal/non-thermal heating and cooling processes in the corona. Part of the obtained deviations 
%could originate from a systematic bias in the emissivities of the strongest IOI lines. In fact, the deviation in 
%ion concentration can further affect the radiative/dielectronic recombination and innershell ionization line strengths,
%directly leading to deviations in the line fluxes. Strictly speaking, one should consider the obtained values in Fig.~\ref{fig:ibal} as an upper limit
%to the atomic uncertainty in ion concentration.

\subsubsection{Uncertainties on line intensities}

Next we address the uncertainties on line emissivities. First, we define the lines of interest (LOIs), which are set as  
a collection of the strongest X-ray lines of the IOIs. For each Fe ion, we pick up the top 200 lines, and exclude those
in the UV range. For the other elements, we do the same but for top 50 lines. These numbers of lines are determined empirically 
to make sure that all the lines visible in the three test spectra are included. The LOIs are selected at the average temperatures
(weighted by the emission measure of each component) of the objects. A full list of the LOIs selected based above criteria
is provided in Table~\ref{tab:longtable}.

To determine the emissivity uncertainty of each LOI, we add for each LOI an emission line with a Gaussian shape to our ultimate spectral model. The line energy is fixed to the energy of the target LOI, the width is determined as described below, and the normalization is directly determined from a spectral fit and represents the difference in flux between the prediction by the ultimate model and the measured spectrum.
A part of the LOIs cannot be resolved as they are blended within the instrumental resolution. In such cases, we
use one Gaussian component for one line blend, and set the Gaussian central wavelength to the average of the 
unresolved LOIs weighted by the emissivities. The line widths of the resolved
Gaussian components are fixed to the thermal plus turbulent broadening of the target LOIs, while for the blended 
lines, the widths are set free. We have tested to let all the line widths free, and found that the best-fit line 
fluxes vary typically by $\leq 20$\%. As shown later in Figure~\ref{fig:ratio_fwhm}, the variation of the line width has a negligible effect
on our conclusion.

The Gaussian components are further corrected for the known astrophysical and instrumental effects, including the 
ionized/neutral absorption, the systematic line shift, and
the residual calibration errors on the effective area as well as on the line spread function as described in \S~\ref{sec:data}. Parameters
of these models are fixed to the values determined from the original fits with model 1. We define model 2 as a co-addition
of model 1, which is now converted to a {\it file} model as to take into account the uncertainties on ion concentration, and the set of Gaussian components. The normalizations of the Gaussian components are the parameters of 
main interest, the positive normalizations account for lines where the model underestimates the data, while the negative
normalizations (or absorption-like components) represent the opposite.

Although the additional Gaussian components with free widths for the blends result in a large number of free parameters
with model 2 (656 for Capella and HR~1099, and 186 for the Perseus cluster), most of the Gaussian components are well 
constrained, or have well defined upper bounds on fluxes (Figs.~\ref{fig:ratio_capella}, \ref{fig:ratio_hr1099}, and \ref{fig:ratio_perseus}). This is because that, at first, all the Gaussian components, including those for the blends, are 
well resolved in the test spectra and fit independently; and secondly, model 2 contains nearly zero degrees of freedom on the astrophysical model (e.g., emission measure, temperature, and abundances) by converting it to a set of {\it file} models. 

%The negative normalizations of the Gaussian components are allowed to account for lines where the model overestimates the data.

Figures~\ref{fig:spec1} and \ref{fig:spec2} plot a comparison of model 1 and model 2 for the Capella HETG spectrum.
It reveals significant discrepancies between model 1 and the observed spectrum in a range of positions, most of those
likely originate from issues in the atomic data that calculate the line fluxes. Most of the discrepancies appear to be solved 
with model 2, which compensates for the mismatches by fitting the extra Gaussian components. As seen in the lower panels 
of the figures, these Gaussian components
contribute up to $\sim 20$\% of the corresponding line fluxes, and vary strongly, in both positive and negative ranges, among different
LOIs or line blends.

For Capella, the C-stat has been improved from 40281.2 with model 1 to 13520.3 with model 2, for an ideal expectation of 6555. 
As shown in Figures~\ref{fig:spec1} and \ref{fig:spec2}, model 2 has fixed the issues in the previous fits 
of \ion{Fe}{XVIII} lines in $14-16$~{\AA}, \ion{Fe}{XIX-XXI} lines in $12-14$~{\AA}, \ion{Fe}{XXII-XXIII} lines around 
11.8~{\AA}, and \ion{Fe}{XVII-XIX} lines in $10-11$~{\AA}. 
Although model 2 has accurately reproduced most of the lines in the spectrum, the overall fit is still formally unacceptable. 
The remaining
discrepancies should be understood as residual errors in instrument calibration, combined with the
minor inadequacies of the atomic data (in particular on wavelengths, see Paper II). Yet there are a few observed weak 
lines, such as
those at 9.64~{\AA}, 9.89~{\AA}, 10.68~{\AA}, 10.86~{\AA}, 16.62~{\AA}, and 17.80~{\AA}, still missing in the present modeling.
Ignoring the spectral bins containing the missing lines could further reduce the C-stat to 10343.

As stated earlier, the present work is based on a systematic comparison between model 1
and model 2 on the known atomic transitions. We will demonstrate later (\S~\ref{sec:wl}) that 
the remaining issues in the fit with model 2, such as the missing weak
transitions at a few places, have a negligible influence on the systematic comparison.  
It would therefore be
valid to consider the Gaussian components, which quantify the difference between model 1 and model 2,
as an approximation to the total uncertainties in the SPEX line modeling with respect to the actual data.
In panel (a) of 
Figure~\ref{fig:ratio_capella}, the model 2 minus model 1 differential flux (i.e., flux of each Gaussian component added to model 2) is plotted against the model 1 line flux for 
each LOI. It shows that the differential fluxes are positively correlated with model line fluxes; a flat distribution 
assuming all Gaussian lines have a common flux can be excluded at $>5\sigma$. This implies that the observed differential
fluxes cannot be fully explained by the systematic uncertainties due to the instrumental calibration, which do not depend 
on the fluxes of the observed lines.

By dividing the Gaussian fluxes by the total fluxes of the corresponding lines, we further obtain the fractional 
uncertainties. As shown in panel (b) of Figure~\ref{fig:ratio_capella}, the uncertainties show anti-correlation with the
line emissivities; for the strong transitions with emissivities $\geq 5\times 10^{41}$ photons s$^{-1}$, 
the fractional uncertainties are found to be around 10\%, while for the weak
lines $\sim 10^{39}$ photons s$^{-1}$, the uncertainties increase to unity or even larger.
In addition, panel (c) of Figure~\ref{fig:ratio_capella} shows that the 
variation of Gaussian line widths has a nearly negligible effect on the observed uncertainty-emissivity relation.

In Figure~\ref{fig:desai} we compare the fractional uncertainties for several \ion{Fe}{XVIII} and \ion{Fe}{XIX} transitions
reported in \citet{desai2005} and those obtained in the our work. It can be found that the present uncertainties
are systematically smaller than those from \citet{desai2005}, for instance, the discrepancies between the model and 
data on 
\ion{Fe}{XVIII} resonance line at 14.208~{\AA} and \ion{Fe}{XIX} line at 13.518~{\AA} have been reduced 
from 30\% and 98\% \citep{desai2005} to the present values of 13\% and 5\%. This means that the current modeling 
of the Capella spectrum, though still far from ideal, has already been improved significantly from the those used in \citet{desai2005} on in particular the atomic database. 

The Gaussian fluxes shown in Figure~\ref{fig:ratio_capella}
can be treated as a combination of statistical and systematic uncertainties. 
By subtracting the statistical uncertainties in quadrature from the total values,
we estimate the systematic uncertainties from the line modeling (Fig.~\ref{fig:ratio_fwhm} left panel).
It can be seen that the  
contributions of statistical uncertainties are minor for most of LOIs, thanks to the high quality of the Capella spectrum. To describe the uncertainty-emissivity relation,
we divide the emissivity range into a number of emissivity bins, and assume for each bin that the distribution of systematic uncertainty follows 
a Gaussian function with zero mean value. As seen in panel (d) of Figure~\ref{fig:ratio_capella}, the total uncertainty 
does show a distribution that can be described by a combination of multiple Gaussian components peaked at zero with different variances. The derived variances 
of the systematic uncertainties are plotted in the right panel of Figure~\ref{fig:ratio_fwhm} as a function of emissivity.
It turns out that the systematic uncertainty-emissivity relation could be approximated by a simple power-law function,
\begin{equation}
\sigma = a \times \left( \frac{I}{10^{41}} \right)^{b} ,
\label{eq:rel}
\end{equation}
where $\sigma$ is the variance of systematic uncertainties, $I$ is line emissivity in unit of photons 
s$^{-1}$, $a$ and $b$ are the free
parameters. The line emissivities are calculated for a standard CIE model with proto-solar abundances \citep{lodders2009} and 
a fixed emission measure of $10^{64}$ m$^{-3}$. As shown in Table~\ref{tab:fitpara}, the parameters $a$ and $b$ are 
found to be 1.020 and -0.563 for Capella. The subsets with positive and negative Gaussian normalizations (black and 
red data points in Fig.~\ref{fig:ratio_capella}) can be described by the same power-law function as the combined set.

%By dividing the Gaussian fluxes by the total fluxes of the corresponding lines, we further obtain the fractional 
%systematic uncertainties. For the strong transitions with emissivities $\geq 5\times 10^{41}$ photons s$^{-1}$, 
%the fractional uncertainties are found to be around 10\%, while for the weak
%lines $\sim 10^{39}$ photons s$^{-1}$, the uncertainties increase to unity or even larger. 
%As show in the right panel of Figure~\ref{fig:ratio_capella}, the observed uncertainty-emissivity relation could
%be described by a simple power-law function, 
%\begin{equation}
%U = a \times \left( \frac{I}{10^{41}} \right)^{b} ,
%\label{eq:rel}
%\end{equation}
%where $U$ is the fractional uncertainty, $I$ is line emissivity in unit of photons s$^{-1}$, $a$ and $b$ are the free
%parameters. The line emissivities are calculated for a standard CIE model with solar abundances \citep{lodders2009} and 
%a fixed emission measure of $10^{64}$ m$^{-3}$. As shown in Table~\ref{tab:fitpara}, the parameters $a$ and $b$ are 
%found to be 0.394 and -0.656 for Capella.
%The positive and negative subsets of the Gaussian fluxes (black and red data points in Fig.~\ref{fig:ratio_capella}, %respectively)
%can be described by the same power-law function as the combined set. In addition, Figure~\ref{fig:ratio_fwhm} shows that the 
%variation of Gaussian line widths has a nearly negligible effect on the analytic fit to the uncertainty-emissivity diagram.

\begin{table}[!htbp]
    \centering
    \caption{Fit parameters and errors of the observed uncertainty-emissivity relations with Eq.~\ref{eq:rel}.}    
    \begin{tabular}{|l|ccc|}
       \hline
         & $T_{\rm peak}$ (keV) & a & b \\
         \hline
        Capella & 0.5 & 1.020 (0.150) & -0.563 (0.067) \\
        HR~1099 & 1.5 & 0.212 (0.084) & -0.762 (0.217) \\
        Perseus & 4.0 & 0.103 (0.183) & -0.946 (0.511) \\
        All     & $-$ & 0.332 (0.058) & -0.623 (0.104) \\
        \hline
    \end{tabular}
    \label{tab:fitpara}
\end{table}

In addition to Capella, we have applied the same exercise to the HR~1099 (Chandra HETG) and the Perseus cluster (Hitomi) spectra. 
The differential fluxes obtained in model 2, the systematic uncertainties, and the variances are plotted against the line emissivity in Figs.~\ref{fig:ratio_hr1099} and \ref{fig:ratio_perseus}. These two objects have higher peak temperatures (1.5~keV for HR~1099 and 4~keV for Perseus) than Capella (0.5~keV). One should also note that the Hitomi spectrum of the Perseus cluster contains only the K-shell lines above 2~keV, whereas the Capella spectrum is dominated by the Fe-L shell lines. Despite of the differences,
the two objects exhibit similar trends in the systematic uncertainty-emissivity diagram as Capella (Figure.~\ref{fig:ratio_all}). This reinforces
the general picture that the strong emission lines are consistently much better modeled than the weak ones.
This picture seems to hold for both the L-shell and the K-shell lines, though the latter is vastly 
overnumbered by the former in the current test.
As shown in Table~\ref{tab:fitpara}, the power-law fits to the HR~1099 and Perseus variances reveal marginal difference from Capella: 
the best-fit relations of HR~1099 and 
Perseus cluster appear to be slightly steeper than Capella, implying for smaller systematic uncertainties in the modeling of  
strong lines for the objects with higher average temperatures. A similar hint can be inferred from the right panel of Figure~\ref{fig:ratio_all}, 
where we compare the systematic uncertainties for the common lines that appear in both Capella and HR~1099 spectra. For the same transition,
the fractional error obtained with Capella (low temperature) is systematically higher than that with HR~1099 (high temperature), indicating that the uncertainty on line flux is likely a temperature-dependent variable rather than a constant. Verifying this possible dependence would require a follow-up study 
with a systematic spectroscopic sample to cover both the L-shell and K-shell emissions. 

\subsubsection{Influence of the weak transitions}
\label{sec:wl}

As described earlier, a few weak emission features in the Capella spectrum are not fully accounted for by the present 
model, indicating that the atomic data for the minor transitions are not yet complete. As the missing line issue 
could also occur to the blends with other lines, it thus becomes vital to evaluate the effect of weak lines on the 
results obtained so far. 

In Figure~\ref{fig:ignoreweak} we compare the original fractional uncertainties of Capella 
with the uncertainties obtained with a modified model, in which the Gaussian components of the known transitions of 
emissivity below $1.5 \times 10^{39}$ photons per second are removed in the fit. Ignoring these components means
many weak lines are wrongly modeled, even so, it seems that they do not alter much the uncertainties of 
the strong transitions. The uncertainty-emissivity relation of the strong transitions remains largely intact in 
the comparison; parameters $a$ and $b$ from Eq.~\ref{eq:rel} change by 6\% and 1\% from the original values in Table~\ref{tab:fitpara}. 
Therefore, our results obtained so far should be robust against the present limitation in atomic data on weak transitions.

\subsection{Dependence on various factors}

In order to understand the origin of the observed uncertainty-emissivity relation, here we examine its possible dependence on 
several variables including the line wavelength, ion species, and dominant line formation processes. As shown in 
Figure~\ref{fig:wavelength}, the total uncertainty on the line flux is plotted as a function of wavelength for each
individual line in the Capella and HR~1099 spectra. 
We find no clear evidence for dependence on line positions; the relative errors seem to be equally distributed across the 
energy band, except for a small group of lines at long wavelengths ($\sim 19$~{\AA}) where the uncertainties are larger
than the average. Figure~\ref{fig:spec2} shows that the present spectral model of Capella does not fully reproduce the observed  continuum between $17-19$~{\AA}, which might partially explain the large line uncertainties. However, this potential bias only 
affects a small
subset of the lines, which is minor to the uncertainty-emissivity relation obtained with the entire set. Therefore, we conclude that
the systematic uncertainties on line fluxes are independent on the line positions.

Next we divide the entire line sample into groups by the ion species. As shown in Figure~\ref{fig:ratio_ion},
the Fe ion groups can be found at different positions in the line emissivity range, which is primarily determined 
by the ion concentration of the source. In general, the distribution of each individual group appears to follow the 
combined distribution
described by the analytic form (Eq.~\ref{eq:rel}). There might be a small number of minor biases in individual groups, such as 
\ion{Fe}{XIX}, which shows a flatter uncertainty-emissivity relation than the average one for HR~1099, though for 
Capella it agrees well with the average relation. We also find
good agreement between the Fe group and the non-Fe group. Therefore, it is likely that the same power-law dependence of 
spectral uncertainties on line emissivity can be applied to most of the ions in the present collisional plasma model.

Finally let us consider the effect of line formation processes. We calculate the contributions to
the upper levels of the LOIs by various processes: collisional excitation, radiative recombination,
dielectronic recombination, and the cascades from these processes to lower levels. The fractional
contributions calculated for Capella are shown in Table~\ref{tab:longtable}. As seen in Figure~\ref{fig:ratio_de},
there is no obvious correlation between the line uncertainties and the collisional excitation contribution
for Capella and HR~1099. The same is found for the dielectronic recombination (though with much less data; Fig.~\ref{fig:ratio_dr}) and the cascade contribution (Fig.~\ref{fig:ratio_allcas}). As for the radiative
recombination (Fig.~\ref{fig:ratio_rr}), we can see a hint for a weak positive correlation against the line 
uncertainty, however, the current significance is rather low, as the data points with large radiative recombination contribution are very sparse. Overall, it suggests that the observed uncertainty-emissivity relation is unlikely
to be fully ascribed to one specific line formation process, but rather caused by the atomic uncertainties in
the theoretical calculations of multiple relevant processes.

It should be noted that some line formation processes might be further affected by the physical properties of stellar 
coronae, for example,
the finite density and electric or magnetic fields. As discussed in, e.g., \citet{mewe1999}, the dielectronic 
recombination rate might be suppressed in high density plasma due to the ionization of doubly excited states, while
it can be enhanced by a factor of $5-10$ by the influence of external electric fields. These two effects might explain 
some individual scatters observed in the Capella and HR~1099 spectra, however, they alone cannot explain the uncertainty-emissivity
relation due to the scarcity of strong dielectronic recombination lines in the observed spectra (Fig.~\ref{fig:ratio_dr}).

\begin{figure*}[!htbp]
\centering
\resizebox{0.9\hsize}{!}{\includegraphics[angle=0]{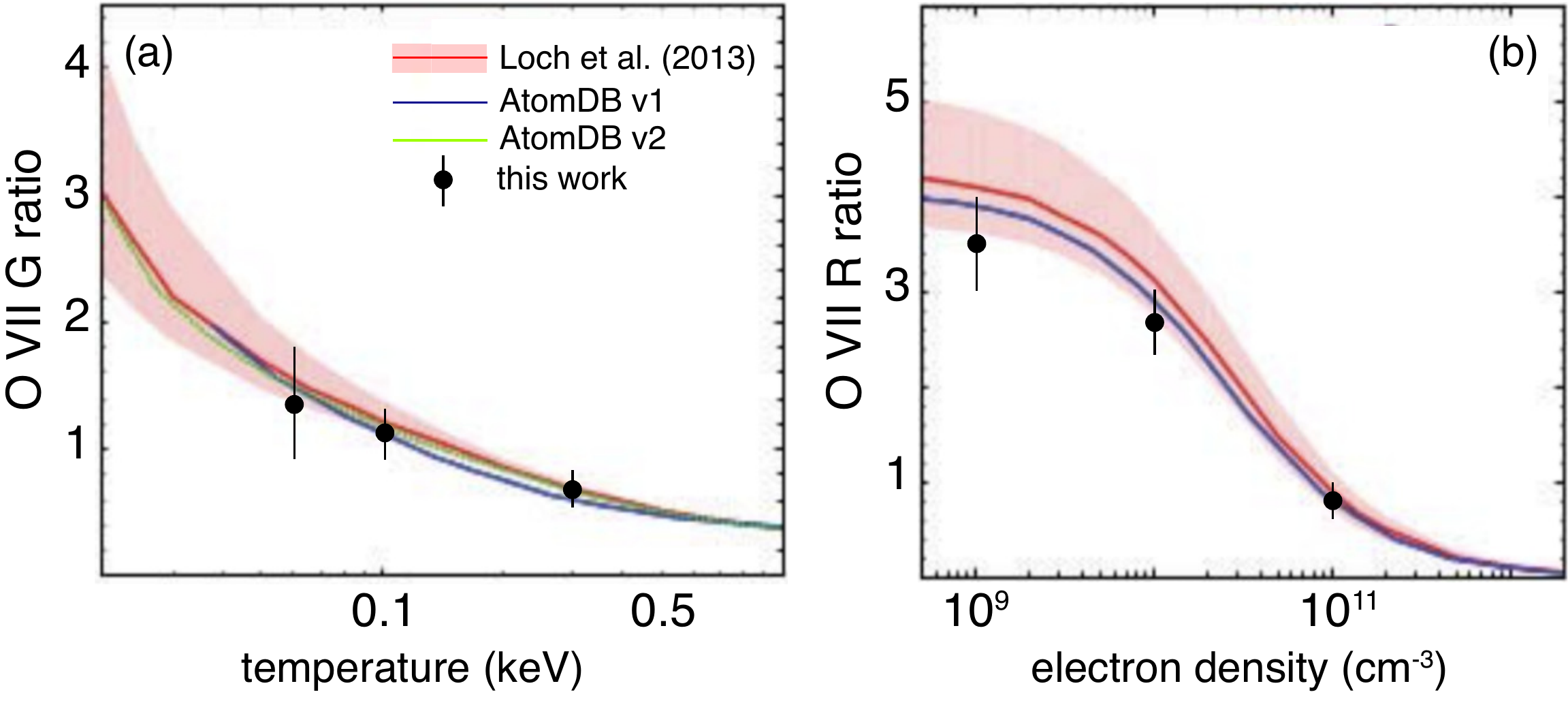}}
\caption{(a) G ratio of \ion{O}{VII} as a function of temperature 
with associated uncertainties calculated for low density plasma. 
(b) R ratio of \ion{O}{VII} as a function of electron density with
an electron temperature of $10^6$~K. The black data with errors show
our results, the red curves and shaded areas represent the peaks and
the associated errors reported in \citet{loch2013}. The blue and green curves show
the results with AtomDB \citep{smith2001, foster2012}. }
\label{fig:compare}
\end{figure*}

\begin{figure*}[!htbp]
\centering
\resizebox{0.9\hsize}{!}{\includegraphics[angle=0]{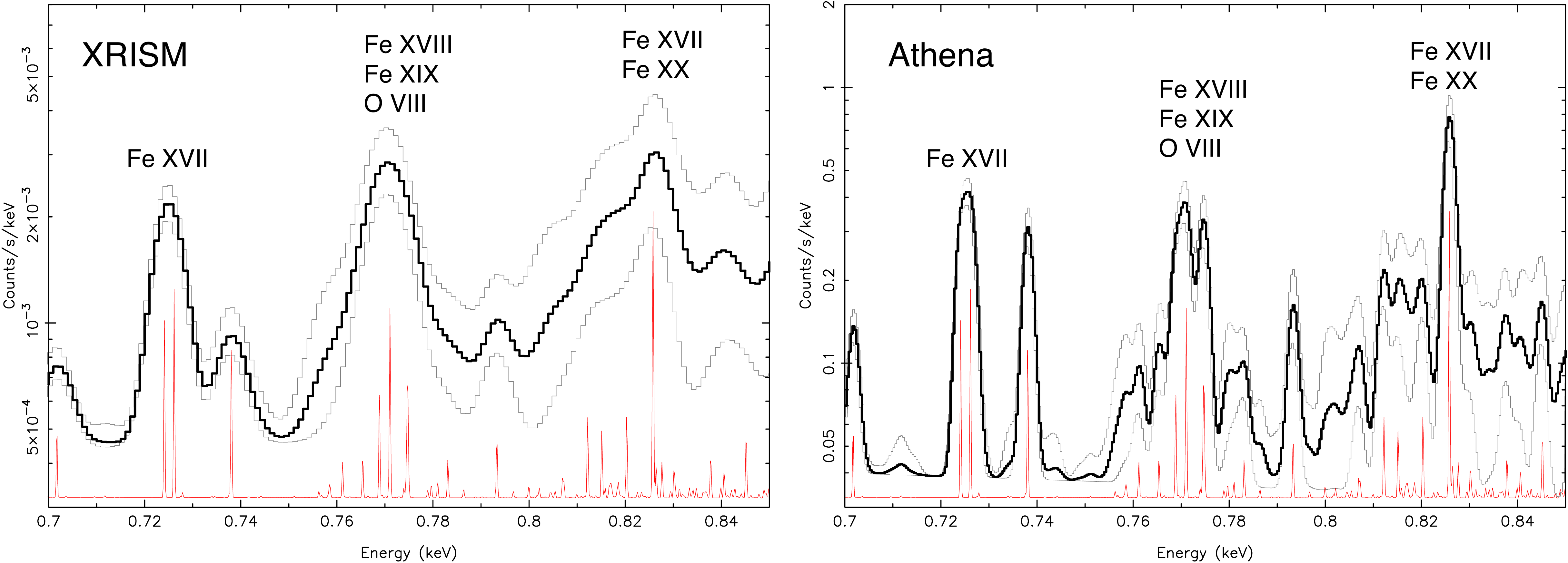}}
\caption{A part of the simulated XRISM (left) and Athena (right) spectra for a 1~keV CIE model, with the associated uncertainties
shown in thin grey curves. The original model is plotted in red.}
\label{fig:simulate}
\end{figure*}

As a summary, we discover a relation between the systematic uncertainties and the line emissivities based
on the high-quality spectra of Capella, HR~1099, and the Perseus cluster. This power-law like relation
holds for spectra with different instruments, objects with different temperature, and lines at different 
positions and from different ions. It cannot be explained by errors in individual line formation process. The observed
relation might describe a universal feature of the state-of-the-art atomic database.

\section{Applications}
\label{sec:application}

In this paper, we present a study of systematic uncertainties in modeling of collisional 
sources using observational data. A key application of the results (e.g., Eq.~\ref{eq:rel}) would be to provide quick and reliable estimate of atomic uncertainties for a range of relevant 
observables, including
in particular the line ratios (e.g., He-like triplet), emission measure, temperature, and abundances. These parameters are the key science outcomes
from the present and future X-ray spectroscopic instruments, their intrinsic uncertainties are vital, albeit so far often missing, 
to our understanding of the astrophysical sources.

\subsection{He$\alpha$ triplet line ratios}
\label{sec:compare}

At first we apply our phenomenological relation (Eq.~\ref{eq:rel}) to evaluate the systematics on the 
He-like line ratios. In Figure~\ref{fig:compare}, we show the uncertainties 
associated with the \ion{O}{VII} He$\alpha$ triplet line ratios compared with the previous results. 
For a low-density plasma, the error on the G ratio is found to be $\sim 35$\% at 0.05~keV and $\sim 5$\% at 0.5~keV. 
The uncertainty on the R ratio with a temperature of $10^6$~K is $\sim 15$\% at an electron density of $1\times10^9$ 
cm$^{-3}$ and diminishes with increasing density. These values in general agree with the uncertainties 
for the same transitions reported in \citet{loch2013}, though the two are derived with very different approaches.
The uncertainties of \citet{loch2013} were obtained from a Monte-Carlo calculation with fundamental atomic constants,
while our values come directly from observations. 

We attempted to do the same comparison with the published results on the lines in ultraviolet \citep{yu2018,dz2019}, however,
the present quality of atomic database in SPEX for these ultraviolet lines is still insufficient to provide 
useful constraint on their emissivities.

\subsection{Atomic uncertainties for XRISM and Athena}
\label{sec:xrism}

Next we apply the obtained line uncertainties to realistic simulations with complicated spectral models, and estimate
the induced errors on the primary model parameters, e.g., plasma temperature, emission measure, 
and elemental abundances. We simulate a set of spectral model by varying all the model line intensities using
the observed uncertainty-emissivity relation (Eq.~\ref{eq:rel} and Table~\ref{tab:fitpara}). For simplicity we
use the average relation with $a=0.332$ and $b=-0.623$. To avoid negative 
or absorption features, the lower boundary of new line emissivities is set to zero. The upper limit on the fractional
uncertainty is set to 100 for the weakest transitions, which is approximately the maximal uncertainty observed in 
the Capella spectrum (Fig.~\ref{fig:ratio_capella}). We test our method on the collisional ionization equilibrium 
spectrum for a set of temperature grids from $0.5-2.0$~keV (Table~\ref{tab:uncert}), which is about the temperature 
range of Capella and HR~1099 where the Fe-L lines dominate the spectrum. We intend to limit our exercise
to relatively low temperatures as the obtained uncertainty-emissivity relation is primarily determined
by the Fe L-shell lines from the observed spectra.

The simulated models are folded with the instrumental responses of XRISM Resolve and 
Athena X-ray Integral Field Unit (X-IFU). XRISM (due to launch in the early 2020s) and Athena (early 2030s) 
are two future X-ray observatories exploring the hot and energetic baryons in the Universe. These two missions will enable 
very well-resolved X-ray 
spectroscopy of various X-ray sources and will push the atomic modeling to
its limit. 
In Figure~\ref{fig:simulate} we illustrate a part of the spectrum, as well as the associated systematic
uncertainty, of a turbulence-free 1~keV CIE model. We run the simulation 1000 times for each 
temperature grid and instrument, fit the randomized data with the original spectral model, 
and summarize in Table~\ref{tab:uncert} the obtained standard deviations of the primary
model parameters: emission measure, temperature, and abundances. For the abundances, we show
the atomic uncertainties on Fe as well as several representative elements in the Fe-L region (O, Ne, Mg, and Ni).
For the latter, the uncertainties originate from the errors on their own emission as well as on 
the blended neighbor Fe-L lines.

\begin{table}[!htbp]
    \centering
    \caption{Fractional systematic uncertainties on model parameters based on the simulation described in \S~\ref{sec:application}.}
    \begin{tabular}{|l|c@{\hskip 0.1in}c@{\hskip 0.1in}c@{\hskip 0.1in}c@{\hskip 0.1in}c@{\hskip 0.1in}c@{\hskip 0.1in}c|}
       \hline
kT & $\sigma_{\rm EM}$ &$\sigma_{\rm kT}$ & $\sigma_{\rm O}$ & $\sigma_{\rm Ne}$ & $\sigma_{\rm Fe}$ & $\sigma_{\rm Mg}$ & $\sigma_{\rm Ni}$ \\
\hline
(keV) & \multicolumn{7}{c|}{XRISM} \\
\hline
0.5   & 0.038  & 0.011  & 0.047  & 0.058  & 0.044  & 0.078  & 0.076 \\
0.75  & 0.063  & 0.007  & 0.067  & 0.081  & 0.072  & 0.066  & 0.092 \\
1.0   & 0.066  & 0.007  & 0.059  & 0.097  & 0.065  & 0.065  & 0.130 \\
1.25  & 0.071  & 0.012  & 0.072  & 0.100  & 0.067  & 0.081  & 0.167 \\
1.5   & 0.060  & 0.021  & 0.060  & 0.098  & 0.067  & 0.093  & 0.133 \\
1.75  & 0.045  & 0.023  & 0.054  & 0.097  & 0.057  & 0.104  & 0.171 \\
2.0   & 0.032  & 0.022  & 0.061  & 0.097  & 0.050  & 0.113  & 0.200 \\
\hline
& \multicolumn{7}{c|}{Athena} \\
\hline
0.5   & 0.033  & 0.011  & 0.029  & 0.047  & 0.043  & 0.067  & 0.050 \\
0.75  & 0.039  & 0.007  & 0.040  & 0.059  & 0.039  & 0.048  & 0.056 \\
1.0   & 0.044  & 0.005  & 0.048  & 0.072  & 0.035  & 0.053  & 0.091 \\
1.25  & 0.051  & 0.008  & 0.065  & 0.077  & 0.038  & 0.071  & 0.112 \\
1.5   & 0.052  & 0.019  & 0.080  & 0.087  & 0.053  & 0.092  & 0.128 \\
1.75  & 0.041  & 0.023  & 0.078  & 0.091  & 0.053  & 0.103  & 0.153 \\
2.0   & 0.031  & 0.022  & 0.070  & 0.092  & 0.049  & 0.111  & 0.175 \\
\hline
    \end{tabular}
    \label{tab:uncert}
\end{table}

As shown in Table~\ref{tab:uncert}, the uncertainties on line flux have propagated into $\sim 3-7$\% 
errors on the emission measure (hence $2-4$\% on gas density), $\sim 1-2$\% on temperature, $\sim 4-7$\% on
O and Fe abundances, $\sim 6-10$\% on Ne and Mg abundances, and $\sim 8-20$\% on Ni abundance, for a XRISM-like
spectrum. Some of these parameters are better constrained with Athena X-IFU, as X-IFU has a significantly better
spectral resolution that helps to de-blend the lines in a crowded complex. The atomic uncertainties are further found to be 
temperature-dependent: the fractional temperature error increases by a factor of three from 1~keV to 2~keV, while the abundance 
errors show peaks around $1.25-1.5$~keV. The different behaviors of temperature and abundance errors show that they might
originate from different sets of lines. Note that the obtained systematic uncertainties likely represent a lower limit, as
(1) the continuum uncertainties, and the errors on the ionization balance calculation, are not yet included; and (2) 
in the present simulation the discrepancies on the line intensities 
are assumed to be fully random, which might not hold in reality.

\section{Conclusion}

We present an observational constraint that could be used to calculate the systematic uncertainties in spectral model of
sources in collisional ionization equilibrium.
Our method is based on statistical properties of the discrepancies between model line fluxes and observed values. The uncertainties
are found to be about 10\% for the strong emission lines, and significantly increase towards low fluxes. The observed
uncertainty-emissivity relation can be approximated by an analytic form, which holds for lines with different wavelengths,
ion species, and formation processes. Applying the observed uncertainties to the simulated XRISM and Athena spectra yields
$4-20$\% systematic errors on the elemental abundances measured from these spectra. In the future this work will be extended
to the other spectral components (continuum and absorption features), and to astrophysical sources in photo-ionization and non-equilibrium ionization status. It should be emphasized that our approach based on observational data can provide absolute
uncertainties of the target atomic constants, however, it cannot be used to illustrate the underlying correlations between the uncertainties of different transitions. Ideally, it will require fundamental theoretical calculations, benchmarked by the
observational results, to derive a full picture of the atomic uncertainties.

The atomic uncertainties estimated using the new approach have been implemented in the {\it aerror} command of 
the SPEX code. 

\begin{acknowledgements}

L.G. is supported by the RIKEN Special Postdoctoral Researcher Program.
SRON is supported financially by NWO, the Netherlands Organization for
Scientific Research.
Work by C. Shah was supported by the Max-Planck-Gesellschaft (MPG), the Deutsche Forschungsgemeinschaft (DFG) Project No. 266229290, and by an appointment to the NASA Postdoctoral Program at the NASA Goddard Space Flight Center, administered by Universities Space Research Association under contract with NASA.
J.M. acknowledges the support from STFC (UK) through the University of Strathclyde UK 
APAP network grant ST/R000743/1.
P. Amaro acknowledges the support from Fundação para a Ciência e a Tecnologia (FCT), Portugal, under Grant No. UID/FIS/04559/2020(LIBPhys).
The research leading to these results has received funding from the European Union’s Horizon 2020 Programme under 
the AHEAD2020 project (grant agreement n. 871158).

\end{acknowledgements}

\bibliographystyle{aa}
\bibliography{main}

\begin{appendix}
\onecolumn
\section{Modeling details of the Capella and HR~1099 spectra}
\label{sec:detail}

\begin{figure*}[!htbp]
\centering
\begin{subfigure}{0.5\textwidth}
\includegraphics[width=\linewidth]{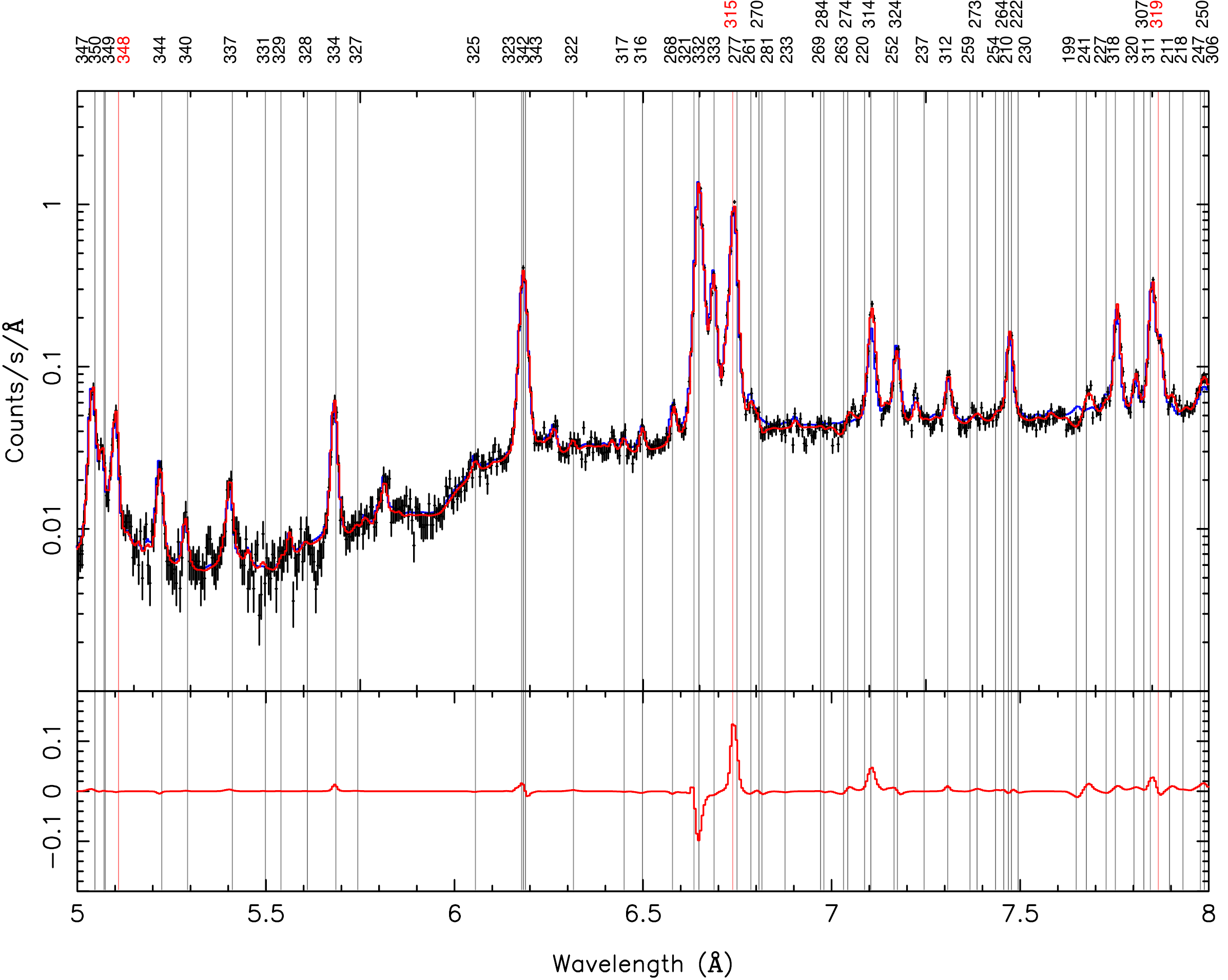}
\end{subfigure}\hspace*{\fill}
\begin{subfigure}{0.5\textwidth}
\includegraphics[width=\linewidth]{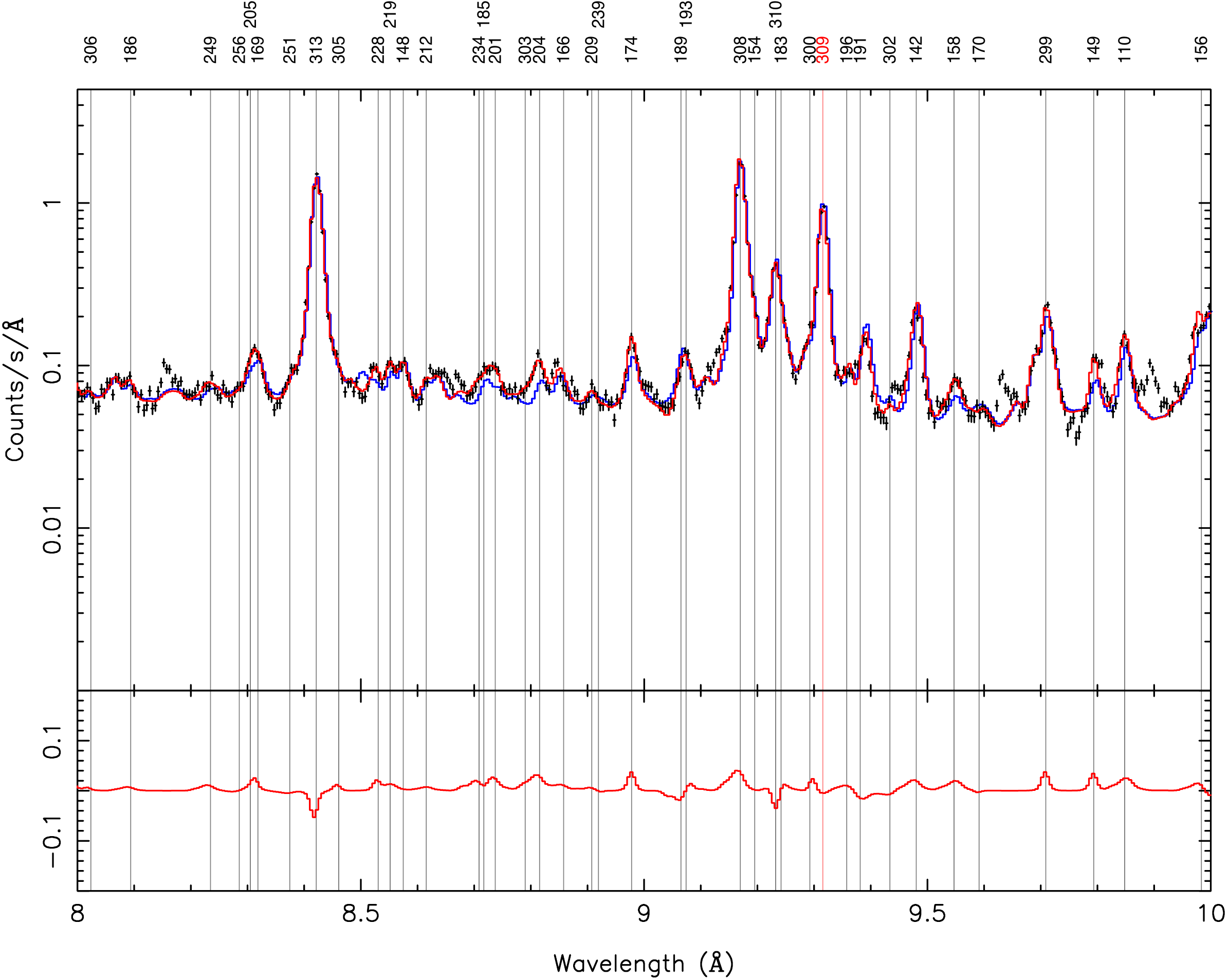}
\end{subfigure}\hspace*{\fill}

\medskip
\begin{subfigure}{0.5\textwidth}
\includegraphics[width=\linewidth]{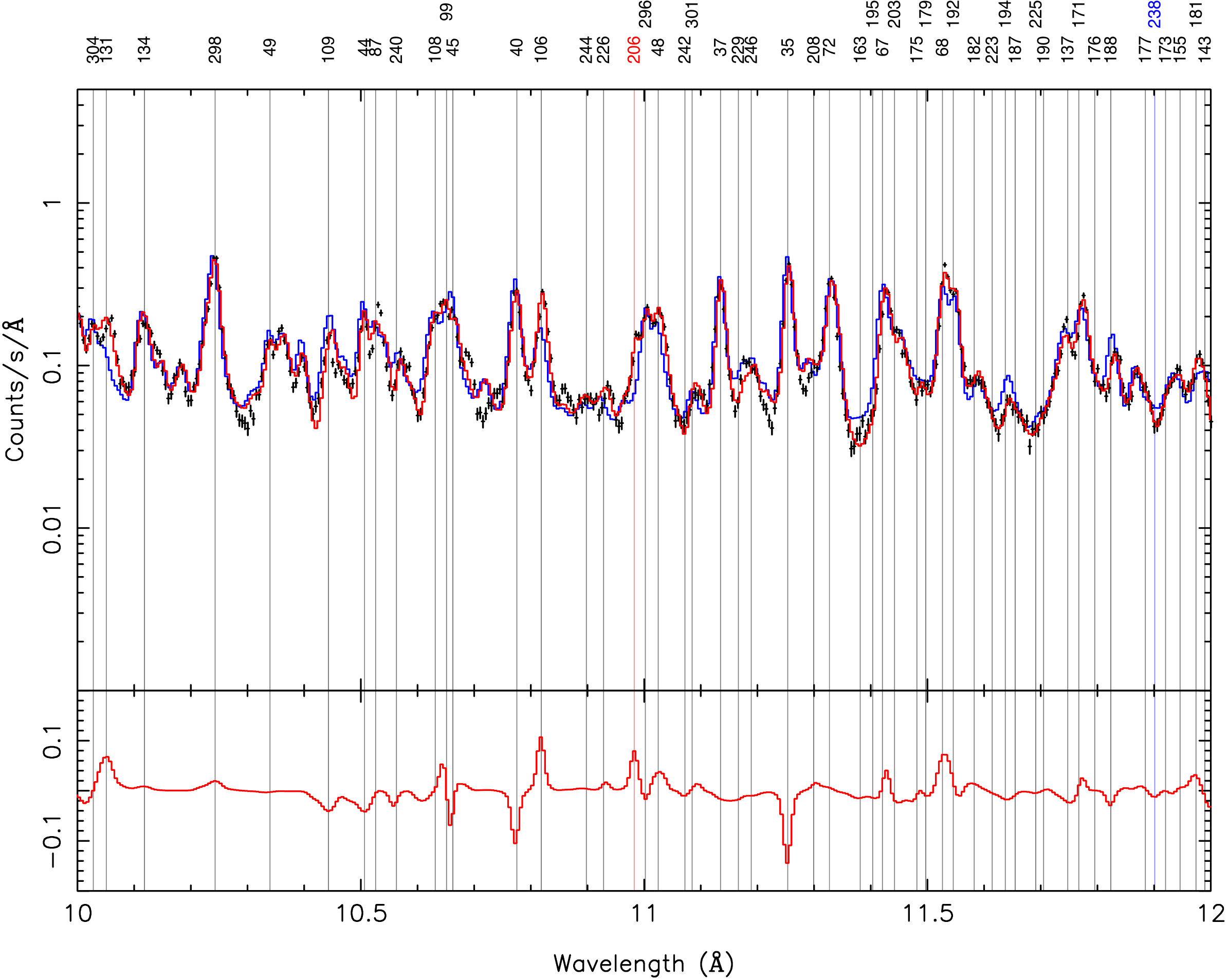}
\end{subfigure}\hspace*{\fill}
\begin{subfigure}{0.5\textwidth}
\includegraphics[width=\linewidth]{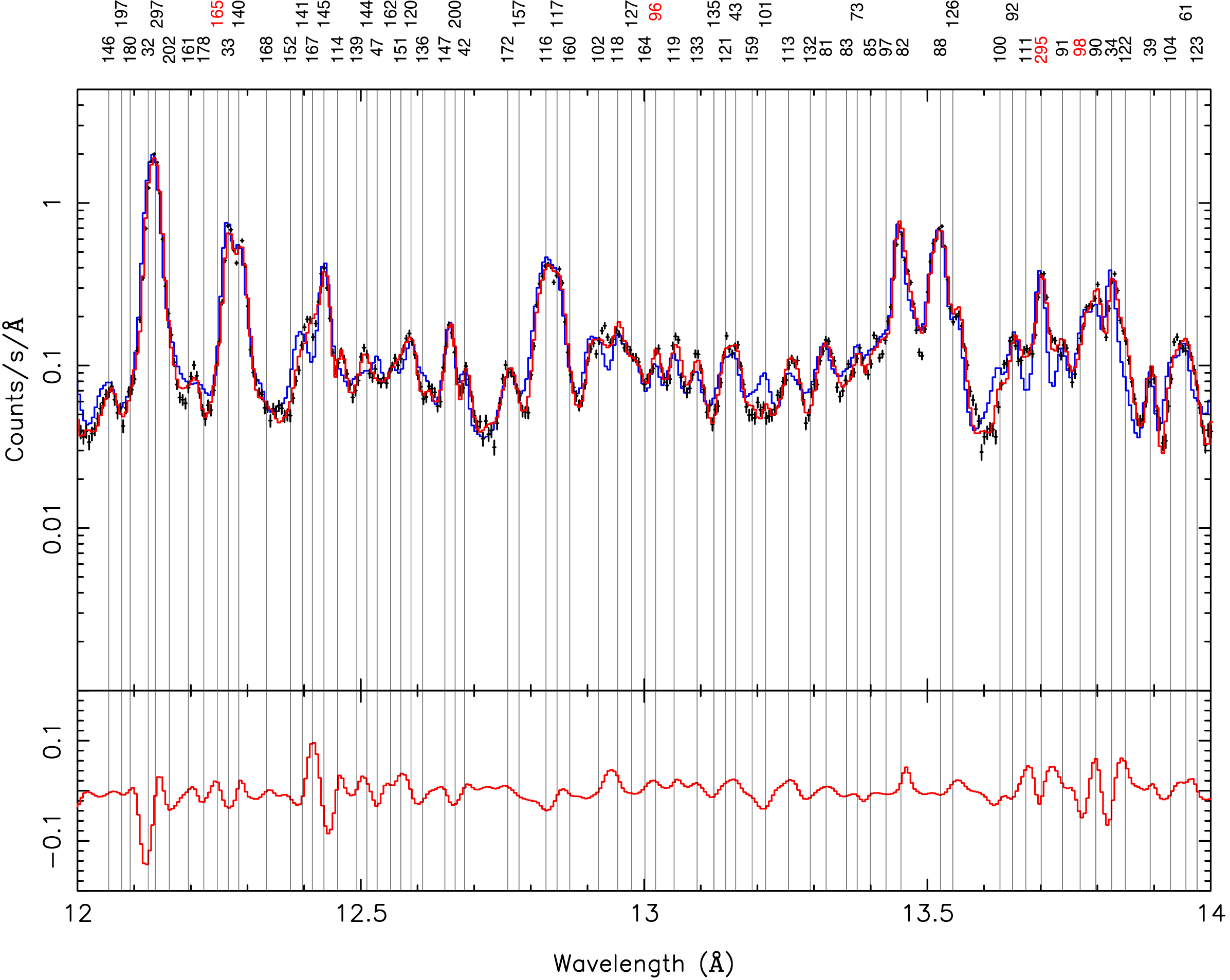}
\end{subfigure}\hspace*{\fill}
\caption{Stacked Chandra grating spectrum of Capella in $5.0-14.0$~{\AA} (wavelength region divided by panels) fit with model 1 (ultimate model from paper II, blue)
and model 2 (file model plus multiple Gaussian components, red). The relative discrepancies between model 1 and model 2 are shown in the lower panel. The selected lines of interest are marked with thin vertical lines. The numbers on the top are the associated
line IDs (see Table~\ref{tab:longtable} for details), the colors of the numbers indicate the dominant line formation process: black $-$ direct excitation; red $-$ radiative cascade; blue $-$ dielectronic recombination. } \label{fig:spec1}
\end{figure*}

\begin{figure*}[!htbp]
\centering
\begin{subfigure}{0.5\textwidth}
\includegraphics[width=\linewidth]{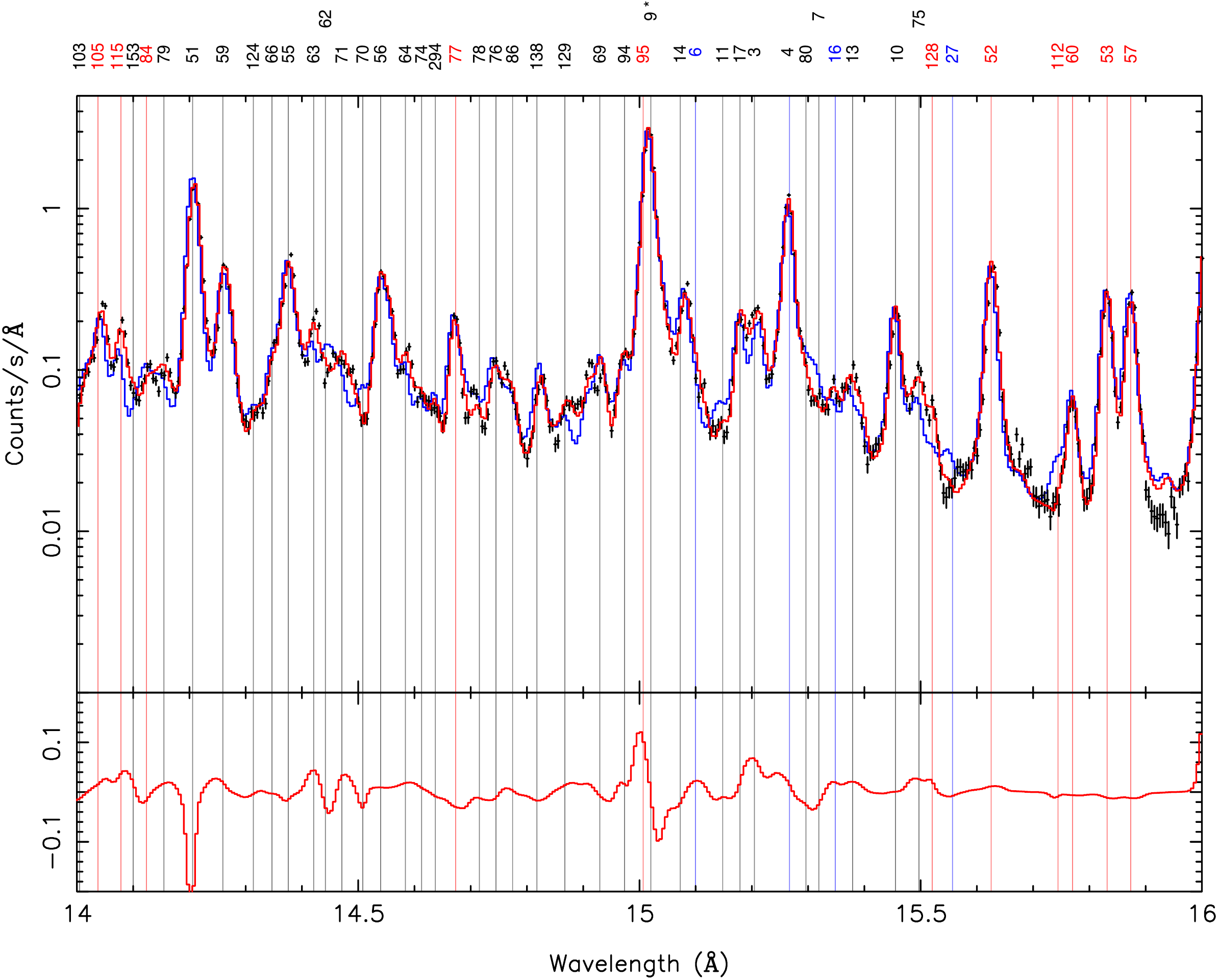}
\end{subfigure}\hspace*{\fill}
\begin{subfigure}{0.5\textwidth}
\includegraphics[width=\linewidth]{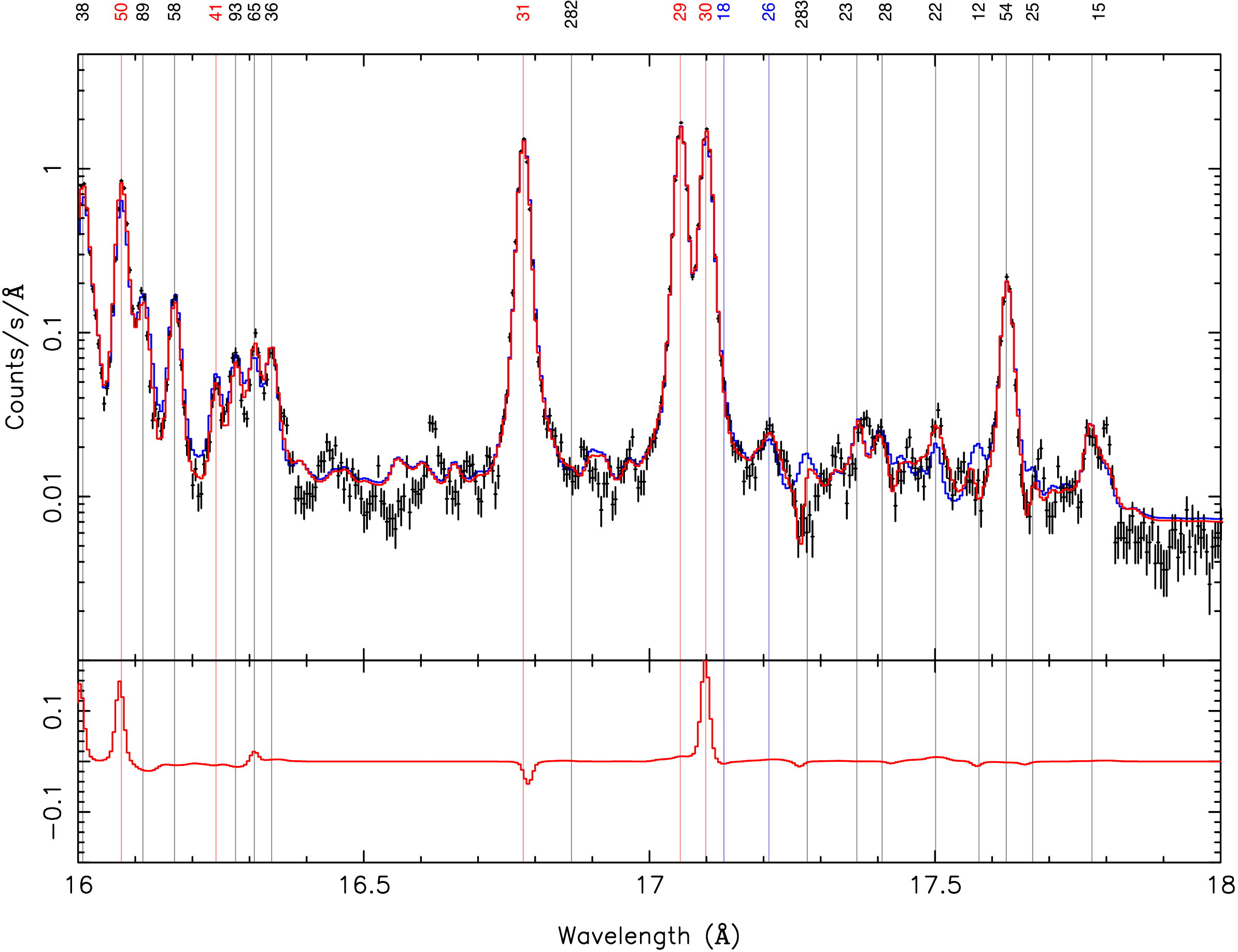}
\end{subfigure}\hspace*{\fill}

\medskip
\begin{subfigure}{0.5\textwidth}
\includegraphics[width=\linewidth]{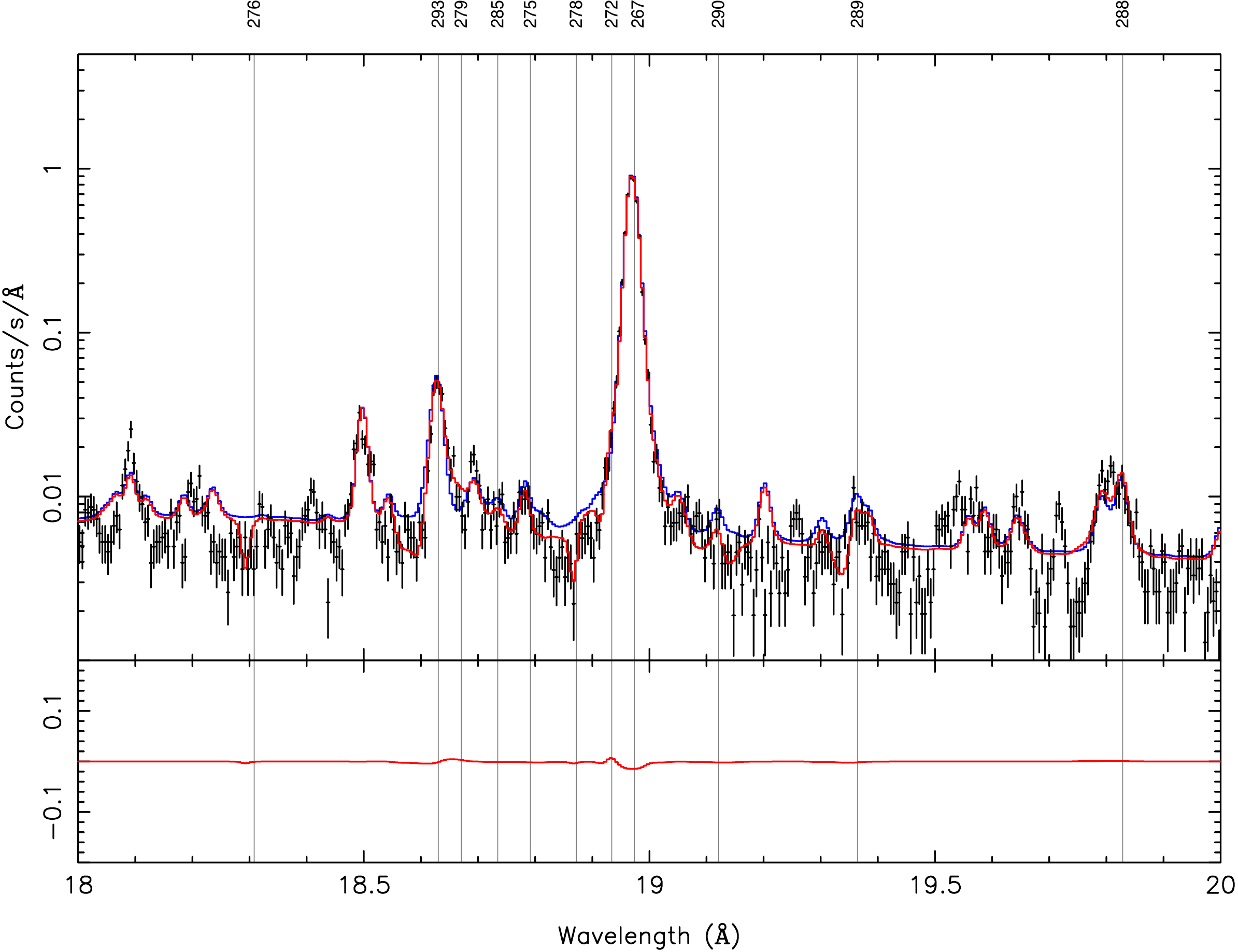}
\end{subfigure}\hspace*{\fill}
\caption{Same as Fig~\ref{fig:spec1} but for $14-20$~{\AA}. } \label{fig:spec2}
\end{figure*}

In Figures~\ref{fig:spec1} and \ref{fig:spec2}, we plot the Chandra HETG data fit with a plasma model and a file plus Gaussian line model (see details in \S~\ref{sec:method}). We also show the ratios between the two models. For each line of interest, 
we list its position, flux, uncertainty, and line formation properties in Table~\ref{tab:longtable}.

\begin{landscape}
\onecolumn
\begin{longtable}{c@{\hskip 0.02in}c@{\hskip 0.02in}c@{\hskip 0.02in}c@{\hskip 0.14in}c@{\hskip 0.14in}c@{\hskip 0.14in}c@{\hskip 0.14in}c@{\hskip 0.14in}c@{\hskip 0.14in}c@{\hskip 0.02in}c@{\hskip 0.02in}c@{\hskip 0.02in}c@{\hskip 0.02in}c@{\hskip 0.02in}c@{\hskip 0.02in}c@{\hskip 0.02in}c@{\hskip 0.02in}c@{\hskip 0.02in}c@{\hskip 0.02in}c}
\caption{\label{tab:longtable} List of lines of interest} \\
\hline\hline
Ion & ID & Wavelength & EX & CEX & RR & CRR & DR & CDR & $I_{\rm Capella}$ & $U_{\rm Capella}$ & $I_{\rm HR1099}$ & $U_{\rm HR1099}$ \\
 & & ($\AA$) & & & & & & & (ph m$^{-2}$ s$^{-1}$) & & (ph m$^{-2}$ s$^{-1}$) & \\
\hline
\endfirsthead
\caption{continued.}\\
\hline\hline
Ion & ID & Wavelength & EX & CEX & RR & CRR & DR & CDR & $I_{\rm Capella}$ & $U_{\rm Capella}$ & $I_{\rm HR1099}$ & $U_{\rm HR1099}$ \\
 & & ($\AA$) & & & & & & & (ph m$^{-2}$ s$^{-1}$) & & (ph m$^{-2}$ s$^{-1}$) & \\
\hline
\endhead
\hline
%\tablefoottext{a}{Emissivities are in units of $10^{-23}$~ph\,m$^3$\,s$^{-1}$.}
\endfoot
\ion{Fe}{XVI} & 1 & 40.1517 & 0.543 & 0.008 & 0.041 & 0.016 & 0.000 & 0.385 & 0.8458 & 0.0448	& 0.2479 & 0.1462 \\
\ion{Fe}{XVI} & 2 & 39.8258 & 0.608 & 0.002 & 0.045 & 0.018 & 0.000 & 0.319 & 0.4057 & 0.1374	& 0.1163 & 0.4795 \\
\ion{Fe}{XVI} & 3 & 15.2008 & 0.000 & 0.000 & 0.000 & 0.000 & 0.934 & 0.066 & 0.4529 & 0.9431	& 0.0616 & 0.5325 \\
\ion{Fe}{XVI} & 4 & 15.2442 & 0.365 & 0.002 & 0.000 & 0.000 & 0.607 & 0.025 & 0.7161 & 0.0719	& 0.0962 & 0.1186 \\
\ion{Fe}{XVI} & 3 & 15.2109 & 0.965 & 0.002 & 0.000 & 0.000 & 0.021 & 0.012 & 1.2664 & 0.9431	& 0.1756 & 0.5325 \\
\ion{Fe}{XVI} & 3 & 15.2140 & 0.000 & 0.000 & 0.000 & 0.000 & 0.973 & 0.027 & 0.3508 & 0.9431	& 0.0456 & 0.5325 \\
\ion{Fe}{XVI} & 5 & 36.7478 & 0.633 & 0.007 & 0.073 & 0.030 & 0.000 & 0.248 & 0.3032 & 21.4497	& 0.0741 & 12.7918 \\
\ion{Fe}{XVI} & 4 & 15.2597 & 0.000 & 0.000 & 0.000 & 0.000 & 0.935 & 0.064 & 0.2516 & 0.0719	& 0.0329 & 0.1186 \\
\ion{Fe}{XVI} & 6 & 15.0960 & 0.182 & 0.001 & 0.000 & 0.000 & 0.760 & 0.056 & 0.3394 & 0.8149	& 0.0452 & 0.2251 \\
\ion{Fe}{XVI} & 7 & 15.3150 & 0.000 & 0.000 & 0.000 & 0.000 & 0.957 & 0.043 & 0.2498 & 1.4674	& 0.0335 & 0.2118 \\
\ion{Fe}{XVI} & 6 & 15.0794 & 0.000 & 0.000 & 0.000 & 0.000 & 0.990 & 0.010 & 0.3070 & 0.8149	& 0.0457 & 0.2251 \\
\ion{Fe}{XVI} & 6 & 15.0752 & 0.000 & 0.000 & 0.000 & 0.000 & 0.997 & 0.003 & 0.3107 & 0.8149	& 0.0463 & 0.2251 \\
\ion{Fe}{XVI} & 8 & 35.1134 & 0.000 & 0.007 & 0.100 & 0.021 & 0.000 & 0.860 & 0.3103 & 2.1108	& 0.0709 & 11.3896 \\
\ion{Fe}{XVI} & 9 & 15.0260 & 0.000 & 0.000 & 0.000 & 0.000 & 0.995 & 0.005 & 0.3035 & 0.1278	& 0.0454 & 0.0230 \\
\ion{Fe}{XVI} & 6 & 15.0916 & 0.000 & 0.000 & 0.000 & 0.000 & 0.991 & 0.009 & 0.2497 & 0.8149	& 0.0371 & 0.2251 \\
\ion{Fe}{XVI} & 9 & 15.0472 & 0.000 & 0.000 & 0.000 & 0.000 & 0.997 & 0.003 & 0.2931 & 0.1278	& 0.0446 & 0.0230 \\
\ion{Fe}{XVI} & 9 & 15.0309 & 0.000 & 0.000 & 0.000 & 0.000 & 0.999 & 0.001 & 0.2940 & 0.1278	& 0.0448 & 0.0230 \\
\ion{Fe}{XVI} & 5 & 36.8018 & 0.662 & 0.022 & 0.071 & 0.021 & 0.000 & 0.216 & 0.1502 & 21.4497	& 0.0365 & 12.7918 \\
\ion{Fe}{XVI} & 1 & 40.2348 & 0.000 & 0.026 & 0.082 & 0.025 & 0.000 & 0.861 & 0.1891 & 0.0448	& 0.0622 & 0.1462 \\
\ion{Fe}{XVI} & 9 & 15.0471 & 0.000 & 0.000 & 0.000 & 0.000 & 0.999 & 0.001 & 0.2363 & 0.1278	& 0.0360 & 0.0230 \\
%\hline
%\caption{ID: line ID (see Figs.~\ref{fig:spec1} and \ref{fig:spec2})}
\end{longtable}

\begin{enumerate}
 \item \noindent Full data table can be found via the link to the machine-readable version.
 \item \noindent ID: line ID (see Figs.~\ref{fig:spec1} and \ref{fig:spec2})  \\
 \item \noindent EX/CEX/RR/CRR/DR/CDR: contributions of upper level population from direct excitation, excitation followed by cascades, radiative recombination, radiative recombination followed by cascades, dielectronic recombination, dielectronic recombination followed by cascades  \\
 \item \noindent $I_{\rm Capella}$, $I_{\rm HR~1099}$: line fluxes from the best-fit models (model 1) \\
 \item \noindent $U_{\rm Capella}$, $U_{\rm HR~1099}$: fractional systematic uncertainties of the lines from the Gaussian fits (model 2)
\end{enumerate}

\end{landscape}

\end{appendix}

\end{document}